\documentstyle[prd,aps,floats]{revtex}
\input{psfig.tex}
\tighten

\newcommand{\beq}{\begin{equation}}
\newcommand{\eeq}{\end{equation}}
\newcommand{\bea}{\begin{eqnarray}}
\newcommand{\eea}{\end{eqnarray}}

\newcommand{\cs}{Compton scattering}

\def\pd {\partial}
\def\nef{n_{\rm e}^{\mbox{\tiny free}}}
\def\nb{n_{\mbox{\tiny B}}}
\def\nh{n_{\mbox{\tiny H}}}
\def\fr{x_{\rm e}}
\def\trec{t_{\rm rec}}

\def\tpi{t_{\rm pi}}
\def\tci{t_{\rm ci}}
\def\Te{T_{\rm e}}
\def\te{t_{\rm e}}
\def\ae{\alpha_{\rm e}}
\def\be{\beta_{\rm e}}

\def\omb{\Omega_{\mbox{\tiny b}}}
\def\thom{\sigma_{\mbox{\tiny T}}}
\def\fcmb{n_{\mbox{\tiny CMB}}}
\def\me{m_{\rm e}}
\def\yff{Y_{\mbox{\tiny ff}}}
\def\yh{Y_{\mbox{\tiny He}}}

\begin{document}
\title{\bf Reionization by active sources and its effects on the cosmic
microwave background} 
\author{Jochen Weller$^1$, Richard A. Battye$^2$ and Andreas Albrecht$^1$\\
$^1$\em Blackett Laboratory, Imperial College, Prince Consort Road,
London SW7 2BZ, U.K. \\{\em and} UC Davis, Dept of Physics, Davis CA 95616, U.S.A
(permanent address).
\\$^2$Department of Applied Mathematics and Theoretical Physics,
University of Cambridge, \\ Silver Street, Cambridge CB3 9EW, U.K.}

\maketitle

\begin{abstract}
We investigate the possible effects of reionization by active sources on the
cosmic microwave background. We concentrate on the sources themselves as  
the origin of reionization, rather than early object formation, introducing an
extra period of heating motivated by the active character of the 
perturbations. Using reasonable parameters, this leads to four possibilities 
depending on the time and duration of the energy input: delayed last
scattering, double last scattering, shifted last scattering and total
reionization. We show that these possibilities are only very weakly
constrained by the limits on spectral distortions from the COBE FIRAS
measurements. We illustrate the effects of these reionization possibilities on
the angular power spectrum of temperature anisotropies and polarization for
simple passive isocurvature models and simple coherent sources, observing the
difference between passive and active models. Finally, we comment on the
implications of this work for more realistic 
active sources, such as causal white noise and topological defect
models.  We show for these models that non-standard ionization
histories can shift the peak in the CMB power to larger angular scales.
\end{abstract}

\section{Introduction}
It is well known that the angular power spectra of temperature anisotropies and
polarization of the  cosmic microwave background (CMB) depend sensitively on
the thermal history of the universe
\cite{Kaiser:84,Vittorio:84,Bond:84,Couchman:86,Naselskii:86,Bond:91,Sugiyama:93a,Hu:95}. 
Around the time of recombination $(z\approx 1100)$, when protons and electrons
recombine into neutral hydrogen, the microscopic physical processes at work are
relatively well understood and the calculation of the photon visibility
function, which feeds into the angular power spectra, is relatively simple, at
least in theories with passive fluctuations, such as inflation.  

However, even in these theories the universe must have become reionized, since
there is no Lyman-$\alpha$ trough in distant quasar spectra (the Gunn-Peterson
test \cite{Gunn:65}). It is thought that this  must be due to virialized
objects, such as protogalaxies, massive stars and quasars, which formed relatively
early in the history of the universe
\cite{Ostriker:90,Tegmark:94,Durrer:1993}. The
microphysics of such processes is less well understood, but photoionization
due to radiative objects can not happen earlier than when these objects have
been created, which is believed to be at redshifts below $z=100$, 
and it is actually thought more likely to have happened much later, after 
$z\approx 30$ \cite {Couchman:86,Ostriker:90,Tegmark:94,Durrer:1993,Fukugita:94,Liddle:95,Arons:72,Shapiro:87,Meiksin:93,Shapiro:94,Haiman:97}.
If this is the case, then the actual observed CMB anisotropies and 
polarization will be a small perturbation on those calculated using the
standard thermal history, just 
including recombination, due to the small optical depth of the time of
reionization. Nonetheless, there are some potentially observable effects,
particularly in the polarization \cite{Zald:96}. 

As always the situation is much less clear in the case of actively generated
perturbations, such as those from topological defects
\cite{zeld:80,vil:81,tursper:88}. In such models, one is forced to try to model
highly non-linear processes from the time of defect formation to the present
day, which is approximately 25 orders of the magnitude in expansion. Even with
the most powerful super-computers available at present, it is difficult to
achieve much more than a factor of 1000 in expansion and hence extrapolations
are necessary. Notwithstanding these difficulties there does seem to be a
consensus at the present as to the predictions of the simplest models, cosmic
strings and textures \cite{PSelTa,ABRa,ABRb,ABRd,ACDKSS,ASWA}, using the
standard thermal history. It appears that flat universe models with critical
matter density ($\Omega_{\rm m}=1$) would require unacceptably large biases
$(\approx 5)$ on 100$h^{-1}$Mpc scales to be consistent with the observed
galaxy distribution \cite{PSelTa,ABRa,ABRb}, although more
exotic defect models may not have this problem \cite{ABRb,Contaldi:98,Vachaspati:99}.  
More acceptable models
can be constructed in an open universe or one dominated by a cosmological
constant \cite{ABRd,ASWA,ACM}. Even, if these models were to be ruled out by
future observations, it is still important to investigate the possibility of
more general active sources as the only credible alternative to inflation. 

The purpose of this paper is to investigate the possible effects of the active
character of such sources for structure formation on the thermal history of the
universe. The basic conceptual difference between active and passive
models, such as inflation, is that the sources are present in the pre- 
and post-recombination plasma. This can create potentially large, local, 
non-linear perturbations, which can accelerate matter causing shock 
heating up to temperatures of a few million Kelvin in the 
baryons\cite{Sornborger:96}.
In order to perform a convincing quantitative treatment of these effects one
would have to incorporate the sources accurately into a full hydrodynamic
simulation which has sufficient resolution to accurately model both the
activation of ionization by the gravitational effect of the sources on the
smallest scales 
and also large enough to model the effects of an expanding universe. Obviously,
the amount of computer resources required for such a simulation would be
prohibitative, and so as a compromize we simply incorporate a gaussian energy
input into the thermal history calculation, which models what we believe would
be the effect of a network of active sources. The {\em effective}
influence of a network is given by a {\em smooth} temperature 
change of the baryons over a certain period of time. A simple way to introduce a smooth 
``jump'' in the temperature is by an errorfunction, so that the heating rate
is a gaussian. This model based approach
allows us to investigate whether there are potentially interesting effects,
before resorting to the more time consuming simulation based approach. This
source has three parameters, the redshift of the maximum energy input, the
amplitude at the maximum and width of the gaussian which models the increase to
and decrease from the maximum. Surprizingly, we find that the limits on the
spectral distortions in the black-body spectrum of the CMB provided by the Far
InfRA-red Spectrometer (FIRAS) instrument on the COsmic Background Explorer
(COBE) satellite, only constrain these parameters very weakly. 

The thermal history calculation yields the so called photon visibility
function \cite{Kaiser:84}, parameterized by time, which acts as a source for
the linear 
Einstein-Boltzmann solver CMBFAST \cite{cmbfast}. This function encodes
statistical information on when the photons which we observe today were last
scattered. For the standard thermal history including just recombination, it
can be modelled as a gaussian centred around $z\approx 1100$ with width $\Delta
z\approx 50$. When we include the energy input there are four interesting
situations which can occur. If the energy input is around or just after recombination then it is
possible to modify the time and length of the last scattering epoch.  If the
energy input occurs once the recombination epoch is ostensibly over, and is
sufficiently short for some of the photons to remain unscattered, then it is
possible to have effectively two surfaces of last scattering, one around the
time of recombination and the other around the time of reionization due to the
energy input. If the energy input is sufficiently late for recombination to be
complete, and long enough for almost all the photons observed today to be
re-scattered at reionization, then it is possible for there to be a single
surface of last scattering at a much lower redshift, effectively shifting the
time of last scattering. Finally, if the period of heating is very late and
long, then the universe remains at least partially ionized for most of the time
after recombination and becomes totally ionized after $z\approx 10$. We shall
describe these four possibilities as delayed, double and shifted last
scattering, and total ionization in the rest of this paper. Of course none of
these possibilities is totally fundamental and just about anything is possible
for a sufficiently complicated source, but they do have some illustrative
value. 

In the next section, we discuss the calculation of the thermal history. First,
we include a detailed review of the standard thermal history  used in most
linear Einstein-Boltzmann solvers. Then we introduce our topological defect
motivated energy source and illustrate the different effects it can have on the
thermal history of the CMB by reference to the photon visibility function. The
effect of this source on the black-body spectrum of the CMB is discussed and we
show that the current limits on spectral distortions would have to be
substantially improved before we could exclude such thermal histories. In  section III, we discuss the effects of these modified thermal histories  on
the angular power spectra of temperature anisotropies and polarization, using
simple analytic arguments to provide qualitative understanding and a linear
Einstein-Boltzmann solver to give quantitative results.  Finally, we discuss the possible implications for more realistic active models such as topological defects and causal white noise models. It should be
noted that we have used natural units ($\hbar=k=c=1$) throughout this paper. 

\section{Thermal history calculations}

\subsection{The standard thermal history}\label{sect-standard}

Originally, the thermal history of the CMB was studied using the Saha equation.
This gave sensible quantitative results, which were subsequently
extended by Peebles and Zel'dovich \cite{Peebles:1968,Zeldovich:69} to include
various corrections due to the 
complexity of recombination to the ground state of hydrogen. These calculations
have been further extended to include more aspects of the underlying Boltzmann
equations for the photons and electrons, and calculations are now at the stage
where further improvements should only lead to about 1\% corrections to the angular
power spectra of temperature anisotropies and polarization \cite{Hu:95}, although even further improvements \cite{Saeger:99} lead to somewhat higher than $1\%$ corrections. In
this section, we review these calculations of the
standard thermal history.   

We quantify the change in the number density of a particular species in terms
of the relevant Boltzmann equations. Strictly, speaking there are seven
equations for the protons, hydrogen, helium, singly ionized helium, doubly
ionized helium, electrons and photons. However, it is sufficient to treat the
helium, both singly and doubly ionized, just using the Saha equation
approximation, since the recombination rate for helium is much faster than the
expansion rate during the relevant epoch \cite{Ma:95}. Also we are really only
interested in the evolution of the photons and electrons since the interaction
of the photons with the relatively heavy baryons is negligible. The number of
photons is much larger than the electrons and therefore their evolution can be
decoupled from the electrons, with this interaction being treated in terms of
the spectral distortions discussed later. Hence, we can model recombination in
terms of the fraction of ionized free electrons  $\fr \equiv \nef/\nh$, where
$\nh = \nb\left(1-\yh\right)$ is the number density of hydrogen nuclei, $\nb$ 
is the number density of baryons and $\yh\approx 0.24$ is the mass fraction of
primordial helium created at Big Bang Nucleosynthesis. In fact, 
\beq
\fr=x_{\rm H}+ {1\over 4}{\yh\over 1-\yh}x_{\rm He}\,,
\eeq
where $x_{\rm H}$ is the  fraction of ionized hydrogen and $x_{\rm He}$ is that
due to helium. Using the Saha equation, one can show that the fraction of ${\rm
He}^{+}$  is about 
$10^{-5}$ for redshifts below $z=2000$ and for ${\rm He}^{++}$ this is even
lower. Hence, $x_{\rm He}$ is always less than $10^{-5}$ and can be included
simply into the calculation using only hydrogen. One slightly odd side-effect
of including helium in this way is that $\fr$ can be greater than one,
although this is only the inclusion of helium modifying the calculation in the
appropriate way; $x_{\rm H}$ is always one or less. The equation which governs
the recombination of protons and electrons into hydrogen is  
\beq
\frac{dx_{\rm H}}{dt} = -\trec^{-1}+\tpi^{-1}+\tci^{-1}\,,
\label{dxe}
\eeq
with $\trec^{-1}$, $\tpi^{-1}$ and $\tci^{-1}$ being the rates for
recombination of hydrogen, photoionization and collisional ionization. In the
standard case, there are no external sources of photoionization and collisional
ionization and,  therefore, we just need to calculate the recombination rate.  

However, to do this we will also need to model the evolution of the temperature
of these distributions. The photon temperature is just redshifted by the
expansion of the universe $T_{\gamma}=T_0(1+z)$ at redshift $z$, where
$T_0=2.728{\rm K}$ is the current temperature of the CMB, while all the other
non-relativistic species remain in thermal equilibrium with each other. One can
derive the evolution of the electron temperature
$T_{\rm e}$ using the first law of thermodynamics \cite{Arons:72},  
\beq
\frac{d\Te}{dt} = -2 \frac{\dot{a}}{a}\Te+\frac{2}{3}
\frac{1}{1-3\yh/4+\left(1-\yh\right)\fr} 
\left( \Gamma - \Lambda \right) - \frac{1-\yh}{1-3\yh/4+\left(1-\yh\right)
\fr}\Te\frac{d\fr}{dt}\,, 
\label{temp}
\eeq
where $a$ is the FRW scale factor, $\Gamma$ is
the heating rate per baryon and $\Lambda$ the cooling rate per baryon. The
first term on the right hand side is cooling due to expansion, the third
term characterizes cooling or heating due to the change in the number
of free particles. The second term is a summation of the heating and cooling
rates of the physical processes involved 
\beq
\Gamma =  \Gamma_{\rm src}\,, \qquad
\Lambda = \Lambda_{\rm rec} + \Lambda_{\mbox{\tiny CMB}}\,,
\eeq
where $\Lambda_{\rm rec}$ is cooling due to recombination,
$\Lambda_{\mbox{\tiny 
CMB}}$ is the Compton cooling from the interaction with the CMB photons and
$\Gamma_{\rm src}$ is any energy input which we might postulate, assumed to be
zero for the standard case. This can be further simplified by realizing that
the recombination cooling due to the loss of kinetic energy from the changing
number of free particles is 
\beq
\Lambda_{\rm rec}={3\over 2}T_{\rm e}(1-Y_{\rm He})\trec^{-1}\,.
\eeq
In the standard case, where we have no source of photoionization or collisional
ionization, one can use (\ref{dxe}) to replace $\trec^{-1}$ with $d\fr/dt$,
which then exactly cancels the last term in (\ref{temp})\footnote{Strictly
speaking the description in the text should include the effects of helium, but
exactly the same cancellation will take place when this is included correctly
for the same physical reasons.}. Physically, we are just cancelling off two
processes which are in equilibrium. Hence, in order to calculate the standard
thermal history we must just compute $\trec^{-1}$ and $\Lambda_{\mbox{\tiny
CMB}}$, and then use a numerical routine to solve the differential equations
for $\fr$ and $T_{\rm e}$, although the stiff nature of these differential
equations does require some caution. 

In order to compute these two quantities we shall now assume a flat, $\Omega_0 =1$, universe with the Hubble parameter 
\beq
H(a)=H_0a^{-2}\left(a+a_{\mbox{\tiny eq}}\right)^{1/2}\,,
\eeq
where $a_0=1$, $a_{\mbox{\tiny eq}} = (1+z_{\mbox{\tiny eq}})^{-1}$ and
$z_{\mbox{\tiny eq}} = 2.40\times 10^4 h^2 (T_0/2.728{\rm K})^{-4}$ is the
redshift of radiation-matter equality, with the Hubble constant parameterized in
the usual way, $H_0 =100 h{\rm km}\,{\rm Mpc}^{-1}\,{\rm sec}^{-1}$. 
The recombination rate that we must calculate here is the `net' rate, taking in
account both the recombination and ionization rate to and from {\em all} states
of the hydrogen atom. Recombination directly to the ground state produces a
Lyman-$\alpha$ photon, which, with high probability, immediately ionizes a
hydrogen atom, either the one which it has come from or one of its close
neighbours. Hence, we do not have to consider recombination to the ground state
with the exception of two possibilities. Firstly, some of the Lyman-$\alpha$
photons may be redshifted out of their resonance line, before they have the
chance to be reabsorbed. Failing that the ground state can only be reached by
the two photon decay: $2s_{1/2} \to 2p_{1/2} + \gamma \to 1s +\gamma$. The net
recombination rate for transitions to and from states above the ground state
and these two possibilities for the ground state is given by
\cite{Peebles:1968,Peebles:1993a} 
\beq
\trec^{-1} = \ae x_{\mbox{\tiny H}} \nh C - \be
\left(1-x_{\mbox{\tiny H}}\right)e^{-3\Delta/4\Te}C\,.
\eeq
This complicated expression requires some explanation. Firstly, and most
simply, we define $\Delta\approx 13.60{\rm eV}$ to be the binding energy of
the ground state.  The rate of recombination to all excited levels is
\cite{Gould:1970}   
\beq
\ae = 2A \left( \frac{ 2 \Te} {\pi m_{\rm e}} \right)^{1/2} 
\frac{\Delta}{\Te} \phi^\prime\left(\frac{\Delta}{\Te}\right) \bar{g}\,,
\eeq
where $A=2^5 3^{-3/2}\alpha^3\pi A_0^2 = 2.105 \times 10^{-22} {\rm cm}^2$
given in terms of the fine structure constant $\alpha=1/137$ and the Bohr
radius $A_0 = 0.529 {\mbox \AA}$, and  $m_{\rm e}$ = 0.511 {\rm MeV} is the
electron mass. 
Also included is a quantum correction\footnote{A more
precise fitting formula for the recombination rate is given in
\cite{Hu:95}. However, the effect on the anisotropy power spectrum, by
including these corrections is less than $2\%$.} for radiative effects known as the Gaunt
factor which is given by $\bar{g}\approx 0.943$ for temperatures below
5000K. The function $\phi^{\prime}(\te)$ comes from summing up the interaction
cross-sections of all excited levels and is given by \cite{Gould:1970,Gould:63}
\beq
\phi^\prime\left(\te\right) = \frac{1}{2} \left( 1.735 - \ln{\te} + \frac{\te}{6}
\right)-\frac{1}{\te} e^{1/\te} E_1(1/\te)\,,			
\eeq
where
\beq
\displaystyle{E_1(x) = \int_{x}^\infty e^{-u}/u\;du}\,,
\eeq
is the exponential integral function and $\te = \Te / \Delta$. The ionization
rate $\be$ is related to the recombination rate by a detailed balance argument
and local thermal equilibrium between all the excited states, and is given by 
\beq
\be = \ae \left( \displaystyle{m_{\rm e} \Te\over 2\pi} \right)^{3/2}
e^{-B_2/\Te}\,, 
\eeq
with $B_2 = \Delta/4$ being the binding energy of the lowest lying excited,
$n=2$, state. 
Finally, the correction due the redshift of Lyman-$\alpha$ photons and the two
photon decay is \cite{Peebles:1968,Peebles:1993a} 
\beq
C=\displaystyle{1+K{\cal D} n_{1s}\over 1+K({\cal D}+\be)n_{1s}}\,,
\eeq
where $K = \lambda_\alpha^3 a/ 8 \pi \dot{a}$, $\lambda_\alpha =
1216 {\rm \AA}$ is the wavelength of the Lyman-$\alpha$ photons, ${\cal{D}} = 8.23 {\rm s}^{-1}$ is the net rate of the
two-photon decay and $n_{1s} \simeq (1- x_{\rm H}) \nh$ the number density of
hydrogen atoms in the $1s$ state. At very late
times the density of the baryons and electrons is low and direct 
recombination to the ground state is possible, since the density 
of the produced Lyman-$\alpha$ photons is then low as well 
\cite{Rees:98}. This leads effectively to $C \to
1$ at low redshifts, where we have choosen $z<100$. The inclusion of
this effect did not change the reionization histories we studied.
 
In order to calculate the Compton cooling rate, one must use the Boltzmann
equation for the photon distribution, which in full generality is
\cite{Komp:1957,Zeld:1969,Sun:1969,Chan:1975,Light:1981}  
\beq
    \frac{dn}{d t}(\nu,t) = \left. \frac{\partial n}{\partial t}
\right|_{\rm cs} +
    \left. \frac{\partial n}{\partial t} \right|_{\rm br} +
    \left. \frac{\partial n}{\partial t} \right|_{\rm dc},
\label{boltphoton}
\eeq
where the subscripts refer to \cs, production of bremsstrahlung (free-free)
photons and double~\cs, respectively and $\nu$ is the frequency of the
photons. We only need concern ourselves with the term due to Compton
scattering, since the other two processes lead to negligible cooling. When
integrated over frequency, this gives the Compton cooling rate for Thomson
scattering of hot electrons off photons in the
plasma\cite{Komp:1957,Zeld:1969,Sun:1969,Durrer:1993}  
\beq
	\begin{array}{lcl}
	\Lambda_{\mbox{\tiny CMB}} &=& \displaystyle\frac{\Te - T_\gamma}{T_\gamma} \left( \frac{\nef
\thom}{\pi^2 m_{\rm e} \nb} \right)\int\limits_0^\infty \omega^4 \fcmb \left(\fcmb +1
\right) \; d\omega \\
	& = & \displaystyle\frac{4 \thom \pi^2}{15 m_{\rm e}} \left( \Te - T_\gamma \right)
T_\gamma^4 \left(1-\yh\right)\fr \\
	&\approx& 6.232 \displaystyle\frac{{\rm eV}}{{\rm s}} \left( \frac{\Te -
T_\gamma}{\Delta} \right) \left( \frac {T_\gamma}{\Delta} \right)^4 \left(1-\yh\right)\fr,
	\end{array}
\eeq
where the Thomson scattering cross section is
$\thom = 6.65 \times 10^{-29} {\rm m}^2$ and $\fcmb$ is the undistorted cosmic microwave background spectrum, $\fcmb^{-1} =  e^{\omega/T_\gamma}-1$.

\begin{figure}
\setlength{\unitlength}{1cm}
\centerline{\hbox{\psfig{file=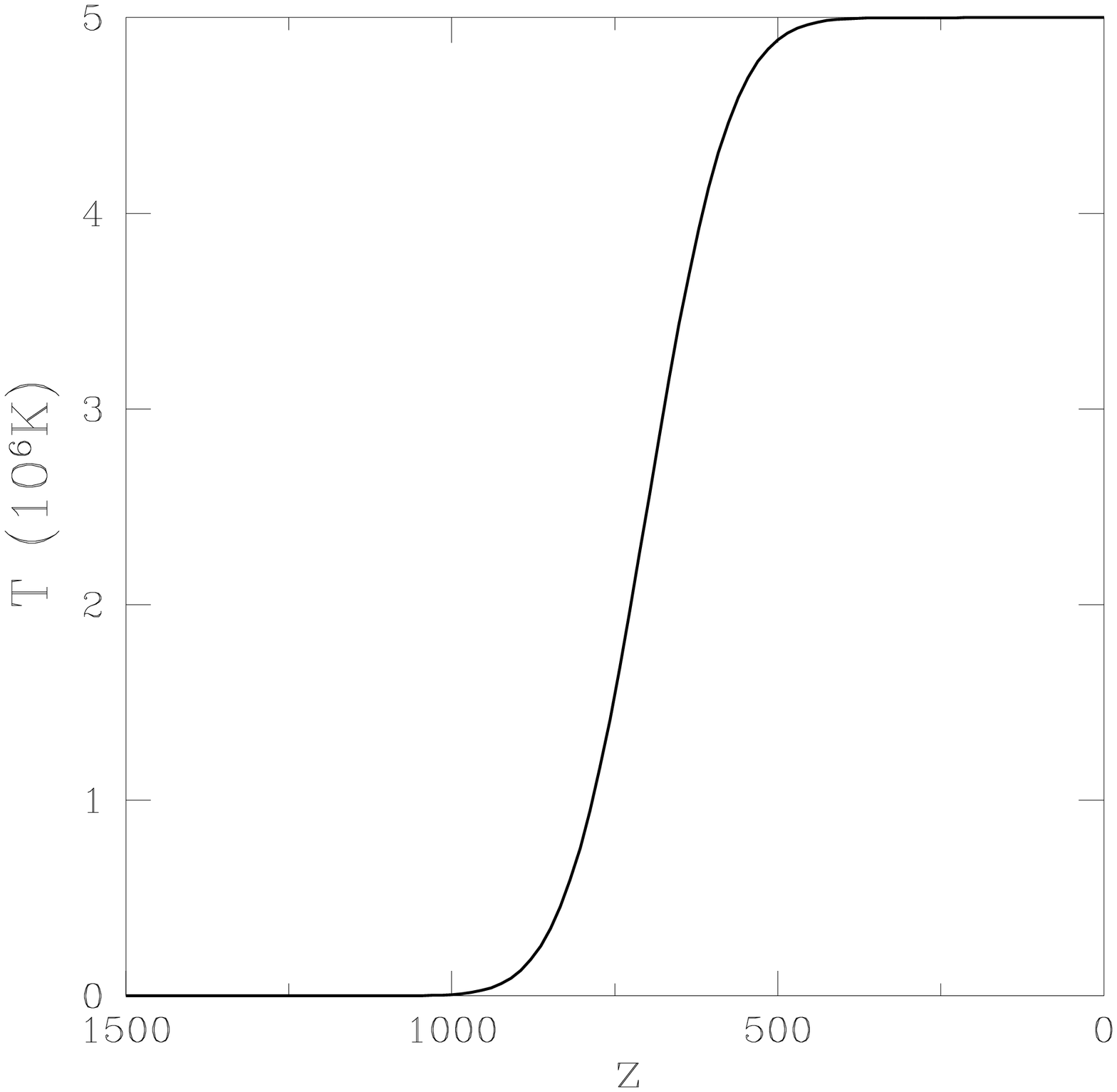,height=6cm,width=7cm}\vspace{4cm}\psfig{file=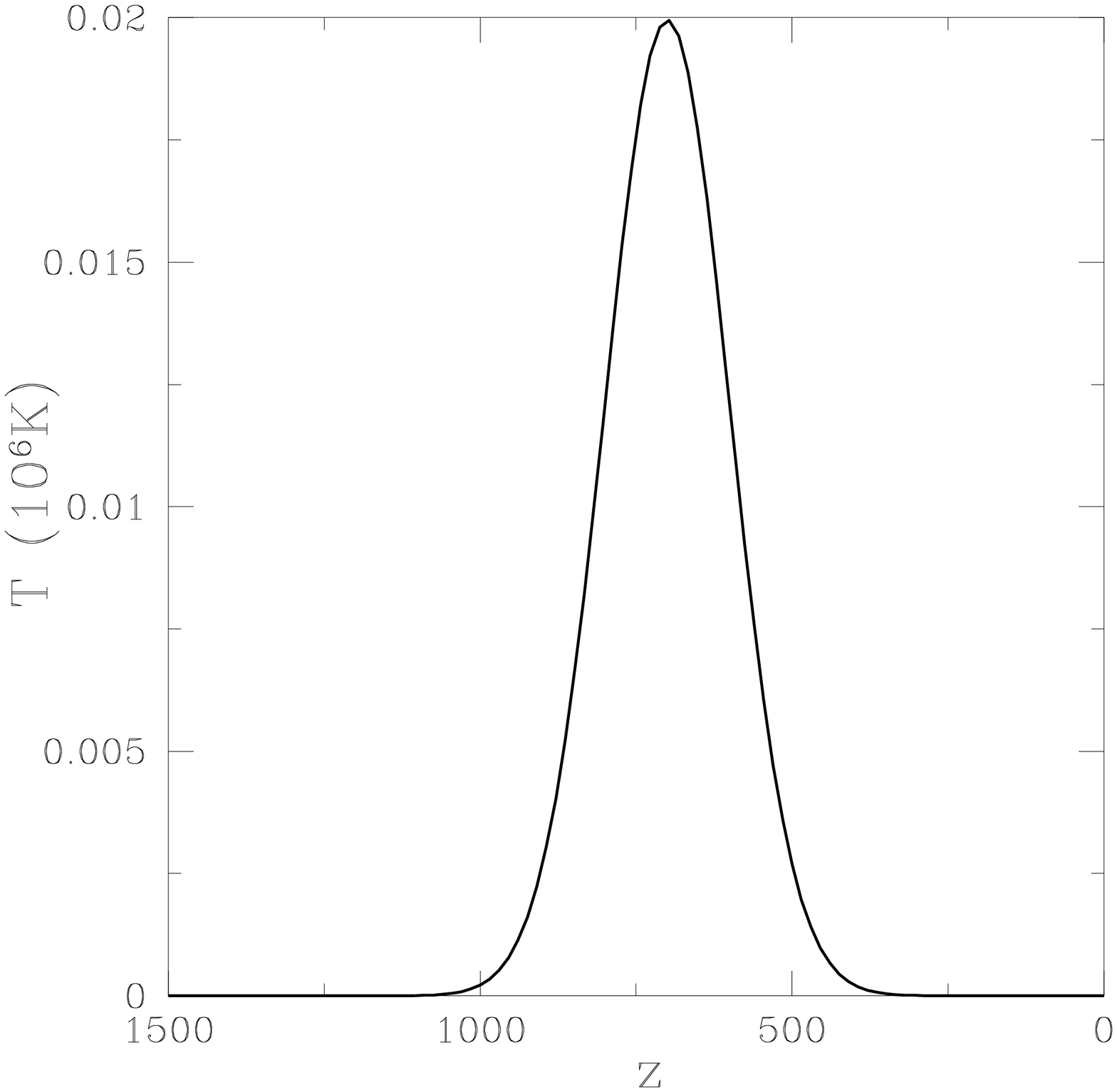,height=6cm,width=7cm}}}
\caption{An example of the energy input. On the left the energy input per free
particle per comoving volume, $q$, as a function of redshift and on the right
the corresponding rate of energy input, $-dq/dz$.} 
\label{fig:input}
\end{figure}

\subsection{The gaussian energy input}\label{sect:input}

In the previous section we reviewed the standard thermal history of the CMB. We
ignored the effects of photoionization, which is usually assumed to be the
source of ionization in the intergalactic medium due to early object formation,
and set to zero the heating due to sources, $\Gamma_{\rm src}$. As already
stated the purpose of this paper is to investigate the possible effects of the
active sources themselves and in this section we introduce a phenomenological
expression for $\Gamma_{\rm src}$, which is intended to model the effects of a
network of topological defects, specifically cosmic strings. In order to do
this we have to make various assumptions which basically allow us to say that
the whole universe becomes ionized in a essentially homogenous way. This is
unlikely to be completely true in a realistic model, since sources are random, but it is required for us to make calculations possible. The effects of
inhomogeneous reionization contribute only to second order in the CMB
anisotropies and will be the subject of future work. 

We assume, therefore, that  the active sources are distributed homogeneously in
the universe and more importantly that the density of these sources is large
enough that they will interact significantly with {\em all} the baryons over a
short timescale.  At early times the thermal velocity of the particles in the
plasma is large and the velocity perturbations will be relatively small, so the
sources will have negligible effect on the baryons. But as the perturbations
grow and the thermal velocities are redshifted by expansion, the sources will
become more significant. In the case of cosmic strings, it has been suggested
\cite{Sornborger:96} that the formation of wakes starts around $z\approx 800$,
which can shock heat the plasma upto temperatures of a few million Kelvin
during subsequent epochs, dependent on the small scale properties of the string
network. 
The increasing kinetic energy of the plasma leads to a partial reionization of
hydrogen, but as the density of active sources decreases, the energy release
in the plasma will also decrease. Therefore, the heating takes place only over
a finite period of time and 
afterwards the temperature of the baryons will remain constant, if, for the
moment,  we neglect the effects of cooling. One could model this effect in many
ways and we have chosen to do this using a smooth function which interpolates
between  temperature of the plasma before the energy input and the temperature
after heating by the sources, once again assuming no cooling. More
specifically,  all the baryons are heated up to a temperature of $T_{\rm heat}$
during a time interval of length $\Delta z=\rho$, centred around $z=\bar z$,
using the energy per free particle per comoving volume  
\beq
q = \frac{T_{\rm
heat}}{2}\left[1+{\sqrt{\pi}\over 2}{\rm erf}\left(
\frac{\sqrt{2}}{\rho}\left(z-\bar{z}\right)\right)\right]\,,
\label{heat}
\eeq
where ${\rm erf}(x)$ is the error function defined, in the standard way, by 
\beq
{\rm erf}(x) = {2\over\sqrt{\pi}}\int_0^x e^{-u^2}\,du\,.
\eeq
Therefore, the rate of energy input has a gaussian shape 
\beq
\Gamma_{\rm src}  =  \displaystyle\frac{3}{2}
\left(1-\frac{3}{4}\yh+\left(1-\yh\right)\fr\right)\frac{dq }{dz}\,\frac{dz}{dt}  =  \displaystyle \frac{3}{2}\sqrt{\frac{2}{\pi}} \left(1-\frac{3}{4}\yh+\left(1-\yh\right)\fr\right)\left(1+z\right)^{5/2}  \frac{H_0T_{\rm
heat}}{\rho}\exp\left[{-\frac{2}{\rho^2}\left(\bar{z}-z\right)^2} \right]\,,
\label{source}
\eeq
where $\rho$ is the width of the Gaussian, $T_{\rm heat}/\rho$ is
proportional to its height and $\bar{z}$ is the position of the peak. The energy
input (\ref{heat}) and the rate (\ref{source}) are plotted in
Fig.~\ref{fig:input} for $\bar z=700$, $\rho=200$ and $T_{\rm
heat}=5.0\times 10^6{\rm K}$. We want to empasize again that our
heating model is just an {\em Ansatz}, which is what we believe a 
reasonable one and allows one to predict the behaviour of the CMB
caused by such a heating process.

Of course, the plasma will never achieve these high temperatures since 
Compton cooling is very efficient for $z > 10$, and in fact we find that it is
difficult for the actual electron temperature to get much above $\Te \approx
5000 \; {\rm K}$. If the effects of the energy input are not significant at
late times, then the temperature of the electrons will drop back to the CMB
temperature, once the heating has stopped. However, it is possible for the
electron temperature to remain around 5000K if the energy input is significant
at low redshift (see, for example, the {\it total ionization} model discussed
below). 

So we now have a source term to add to the standard thermal history which is
motivated by a network of evolving topological defects. In the standard thermal
history we made a number of assumptions, which need to be re-examined in the
presence of this source. Firstly, there is still no source of photoionization,
since ionization by free-free emissions is negligible for our shock motivated
source. The addition of the source can also lead to collisional ionization, but
we have explicitly checked that this is negligible. The main reason
for this is that the densities are not high enough and the fact that 
$\Delta\approx 13.60{\rm eV}\approx 158000 {\rm K}$ is 
much higher than the temperature of the plasma achieved by the thermal input.
Similarly 
we have checked that collisional ionization and excitation do not play a
significant role in the cooling process since they are much weaker than Compton
cooling. 

Hence, we conclude that to model the effects of a topological defect network on
the thermal history of the CMB, one can just modify the standard calculation by
the inclusion of the source term (\ref{source}) bearing in
mind the uncertainties of the {\em Ansatz}. One might wonder how the
ionization occurs, since we have explained above that the usual physical
processes - photoionization and collisional ionization - are negligible. The
basic mechanism is by modifying the rates for recombination and ionization,
$\ae$ and $\be$ respectively, creating a shift in the balance between atomic
hydrogen and free protons and electrons. More specifically, the increase in
$T_{\rm e}$ creates a significant modification to $\ae$ (the probability that an
electron is captured by a proton decreases with increasing $\Te$) and a slightly smaller
effect in $\be$, reducing $\trec^{-1}$, hence shifting the balance towards free
protons and electrons. In other words, ionization dominates over
recombination as long as the matter temperature is large.

\begin{figure}
\centerline{\psfig{file=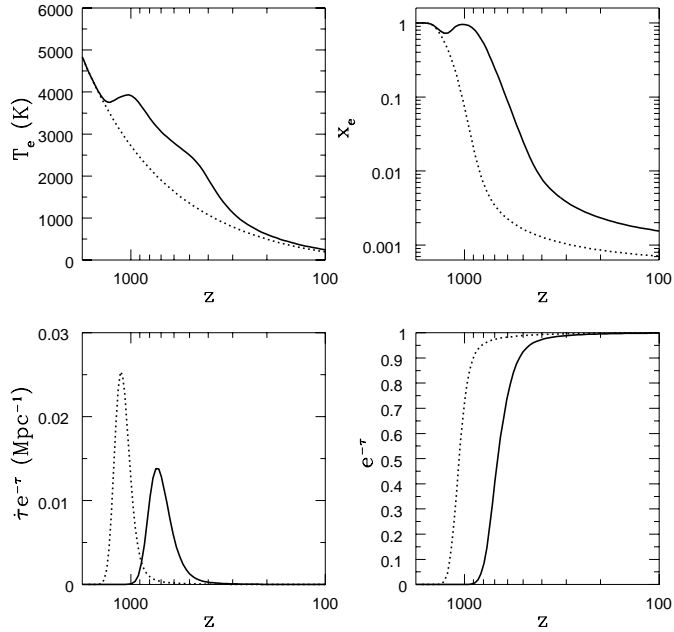,width=3.6in}}
\caption{The effects of {\it delayed} last scattering on the thermal history
and photon visibility functions. We have used ${\bar z}=1000$, $\rho=350$ and
$T_{\rm heat}=1.3\times 10^{8}{\rm K}$. The plots are : (top,left) the electron
temperature $\Te$, (top,right) the fraction of ionized electrons $\fr$,
(bottom,left) the last scattering visibility function, (bottom,right) the
cumulative visibility function, all plotted against redshift $z$. On each plot
there are two curves, the modified (solid line) and standard (dotted line)
thermal histories.} 
\label{fig:delayed}
\end{figure}

\begin{figure}
\centerline{\psfig{file=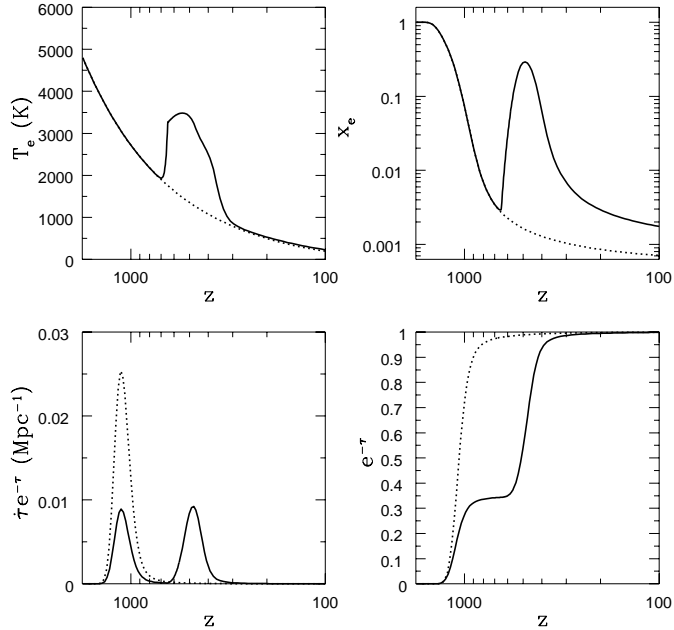,width=3.6in}}
\caption{The effects of {\it double} last scattering using ${\bar z}=500$, $\rho=100$ and $T_{\rm heat}=10^{7}{\rm K}$. The plots are arranged in the same order as Fig.~2.} 
\label{fig:double}
\end{figure}

\begin{figure}
\centerline{\psfig{file=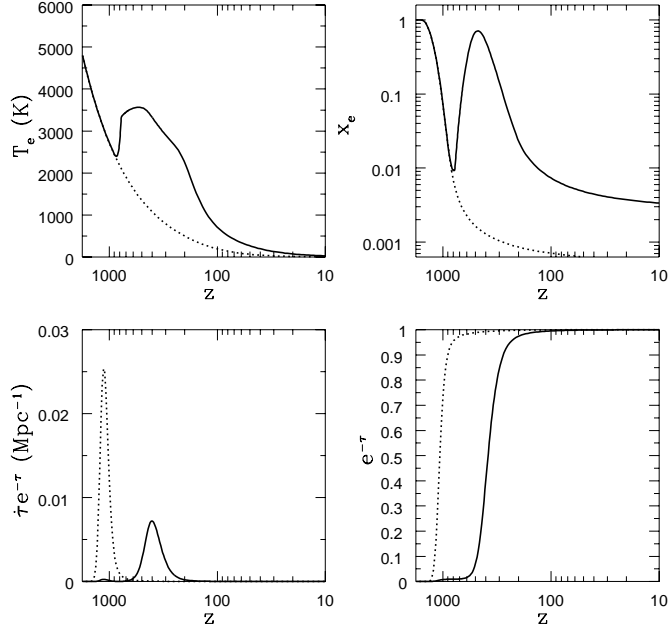,width=3.6in}}
\caption{The effects of {\it shifted} last scattering using ${\bar z}=500$,
$\rho=200$ and $T_{\rm heat}=4\times 10^{7}{\rm K}$. The plots are arranged in
the same order as Fig.~2.}  
\label{fig:shifted}
\end{figure}

\begin{figure}
\centerline{\psfig{file=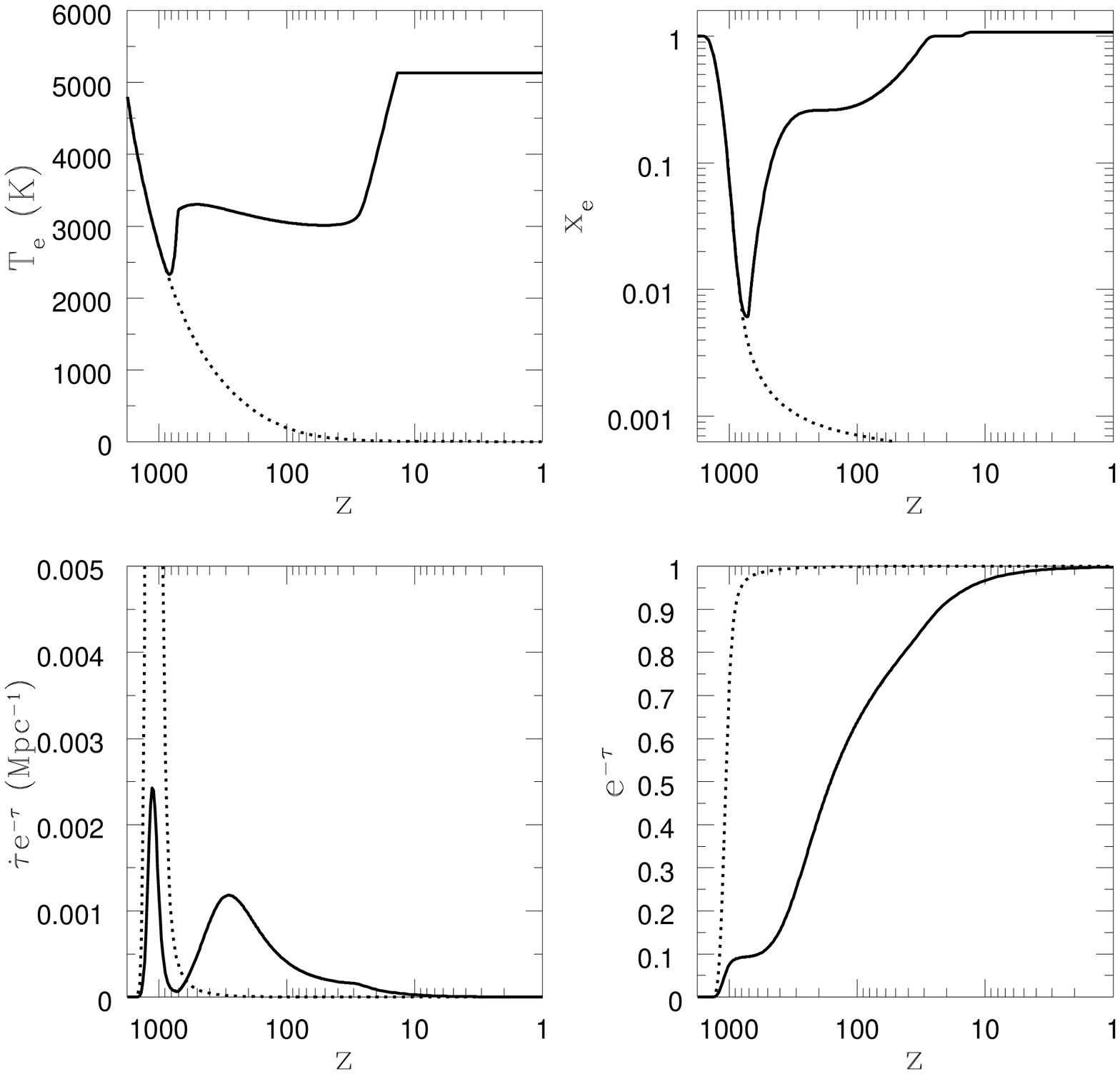,width=3.6in}}
\caption{The effects of {\it total ionization} using ${\bar z}=350$, $\rho=300$
and $T_{\rm heat}=1.5\times 10^{7}{\rm K}$. The plots are arranged in the same
order as Fig.~2.}  
\label{fig:total}
\end{figure}

We have investigated the effects of using this source for a wide range of
parameters and found that there are four cases which can illustrate interesting
effects. We call them {\it delayed}, {\it double} and {\it shifted} last
scattering, and {\it total ionization} and an example of each is discussed
below. Before doing this we should discuss what we shall use to quantify their
effect on the microwave background. The differential optical depth due to
Thomson scattering is $\dot\tau =\fr\nh\sigma_{\rm T}a$ and hence the optical
depth of any particular conformal time $\eta$ is 
\beq
\tau(\eta)=\int_{\eta}^{\eta_0}\dot\tau(\eta^{\prime})d\eta^{\prime}\,,
\eeq
where $\eta_0$ is the conformal time of the present day. From these two
quantities we can construct two photon visibility functions \cite{Kaiser:84},
the last scattering visibility function  
\beq
g(\eta)=\dot\tau(\eta)e^{-\tau(\eta)}\,,
\eeq
and the cumulative visibility function
\beq
h(\eta)=e^{-\tau(\eta)}\,,
\eeq 
which encode statistical information about when the CMB photons observed today
were last scattered or influenced by objects, such as topological defects,
along the line of sight. We have already commented that in the standard thermal
history the last scattering visibility function is a gaussian of width $\Delta
z\approx 50$ centred around $z\approx 1100$. The cumulative visibility
function is even simpler, being effectively zero for $z>1100$ and increasing
sharply to one for $z<1100$ in the standard case. This reflects our inability
to see directly any objects beyond the surface of last scattering since the
universe is opaque. It also has a simple statistical interpretation; for any
given conformal time $\eta$, then $h(\eta)$ is the probability that a given
photon was scattered before $\eta$ and $1-h(\eta)$ is the probability that it
was scattered after $\eta$. We shall see that these functions occur naturally in
the calculation of CMB anisotropies and hence it is important for us to
understand the effects of our energy input on them. 

\smallskip
{\noindent\it Delayed last scattering} - If the energy input is significant
during or close to the epoch of recombination then it is possible to delay
standard recombination. An example of this is $\bar z = 1000$, $\rho = 350$ and
$T_{\rm heat}=1.3\times 10^{8}{\rm K}$, whose effects on the thermal history and
photon visibility functions are illustrated in Fig.~\ref{fig:delayed}. We see
that the temperature of the electrons deviates away from that of the photons
around $z\approx 1500$ and only manages to get back down to be close to the
photon temperature around $z\approx 100$. This allows a significant delay in
recombination process with the ionization fraction remaining greater than
$70\%$ until around $z\approx 700$. The last scattering visibility function is
still similar to a gaussian, but it peaks around $z\approx 700$, as opposed to
$z\approx 1100$, with a larger width and consequentially smaller amplitude. A
corresponding change in the cumulative visibility function is also observed,
with the transition from zero to one happening around $z\approx 700$. 

\smallskip
{\noindent\it Double last scattering} - If the energy input is slightly later,
then it is possible for recombination to take place in the standard way, with
$\fr$ decreasing almost to zero before increasing once again as the electron
temperature begins to rise due to the energy input. If the period of heating is
short enough for a substantial fraction of the photons which were originally
last scattered at the standard time to remain unscattered during the reionized
epoch, then it is possible to have effectively two surfaces of last
scattering. This can be achieved, for example, if $\bar z=500$, $\rho = 100$
and $T_{\rm heat}=10^{7}{\rm K}$, and it effects are compared to the standard
case in Fig.~\ref{fig:double}. As for {\it delayed} last scattering the
electron temperature deviates significantly from that photon temperature, but
now it takes place over a relatively short timescale, between $z\approx 700$
and $z\approx 300$. The first epoch of recombination is effectively over by
around the time at which $\Te$ begins to increase, but this increase leads to
ionization which reaches a maximum of around $30\%$ by $z\approx 450$, before
decreasing again. The last scattering visibility function can be approximated
by two gaussians one centred around $z\approx 1100$ and the other around
$z\approx 500$. The cumulative visibility function has two steps, from zero
to 0.3, and then from 0.3 to one. This tells us that $30\%$ of the photons
observed today  were last scattered around the time of standard recombination,
while the other $70\%$ were last scattered during the epoch of reionization. 

\smallskip
{\noindent\it Shifted last scattering} - If the energy input takes place after
the standard recombination has taken place, as for {\it double} last
scattering, but the duration for which it takes place is much longer, then most
of the CMB photons will be re-scattered during the epoch of reionization and
the photon visibility function will be almost zero around the time of standard
recombination. Although, this is physically distinct from {\it delayed} last
scattering, the photon visibility functions will be very similar apart from a
shift to lower redshift, and hence we call this {\it shifted} last
scattering. Using  $\bar z = 500$, $\rho = 200$ and $T_{\rm heat}=4\times
10^{7}{\rm K}$, one can shift the surface of last scattering to lower redshift,
as illustrated in Fig.~\ref{fig:shifted}. In this case, the electron
temperature is above that of the photons from around $z\approx 800$ until very
close to the present day $(z\approx 10)$, with the fraction of free electrons
being larger than $10\%$ between $z\approx 700$ and $z\approx 100$. The last
scattering visibility function is now approximately a single gaussian centred
around $z\approx 500$, with a much larger width than in the standard case. The
cumulative visibility function can be approximated by a single step at
$z\approx 500$, re-enforcing the fact that almost all the photons were last
scattered during the epoch of reionization. 

\smallskip
{\noindent\it Total ionization} - If the energy input is very late and over a
substantial period then it is very difficult for the universe to become neutral
to any great degree between the epoch of standard recombination and the present
day. In this case, the photon visibility function will be relatively small
since a large fraction of the CMB photons observed today have not stopped
scattering through the whole history of the universe and effectively there is
no well defined concept of last scattering. One interesting side effect of this
is that the universe becomes totally ionized at around  $z\approx 10$ and the
electron temperature does not come back down to the photon temperature,
remaining at around 5000K. An example of this type of thermal history is given
in Fig.~\ref{fig:total} using $\bar z=350$, $\rho=300$ and $T_{\rm
heat}=1.5\times 10^{7}{\rm K}$. The electron temperature is much higher than the
photon temperature for $z<700$, where the energy input becomes significant,
and remains around 3000K until around $z\approx 20$. At this point the effects
of Compton cooling reduce substantially and the electron temperature increases
to be greater than $5000{\rm K}$ for $z<10$. There is a corresponding two step increase in
the fraction of free electrons. The last scattering visibility function is much
smaller than in all the other cases, but still has a peak around the time of
standard recombination and a broader peak across a wide range of redshifts. The
cumulative visibility function increases from zero to about 0.1 around the
epoch of standard recombination and then increases very slowly to one between
$z\approx 500$ and $z\approx 1$. Hence, only $10\%$ of the photons observed
today were last scattered during the epoch of standard recombination and the
other $90\%$ were scattered at some point after $z\approx 500$. The amount of
energy per baryon injected into the universe is 
given by $E/N_{\rm b}=3/2T_{\rm heat}$, which gives in the case of
{\em delayed last scattering} $16.8\,{\rm keV}$ per baryon. After the baryons
reach equilibrium with the CMB this energy input is recognizeable as
spectral distortion of the Planckian spectrum which is discussed in
the next subsection. The resulting distortions should be small since there are
about $10^9$ photons per baryon.

\smallskip
We should emphasize that none of these possibilities is fundamental and the
effects described above will often be superposed in a non-trivial way. Also
only the {\it total ionization} model achieves substantial ionization at late
times and hence some other mechanism, probably photoionization, would be
required to pass the Gunn-Peterson test. 

\smallskip
Although the mechanisms of heating and
reionization discussed in this paper are entirely different, 
the results can look very similar to reionization due to the presence of light 
supersymmetric particles (e.g.~a light photino or higgsino) 
\cite{Salati:84,Asselin:88}. These light inos decay into UV-photons which
subsequently ionize the matter. Dependent on the lifetime of light inos the 
(re-)ionization history is comparable to the one
discussed in the present paper. The remaining energy of the ionizing photons
can heat the matter to much larger temperatures as in the cases discussed here
\cite{Asselin:88}. Other mechanism of reionization which have been
studied in the past are reionization by decaying massive neutrinos
\cite{Sciama:82a,Sciama:82b,Scott:91} or the evaporation of primordial
black holes \cite{Naselskii:86,Naselskii:87}. It should be noted that these 
mechanisms can also lead to reionization histories similar to the ones
presented in this paper.

\subsection{Spectral distortions}

We have already noted that the number of photons is so much larger than the
number of baryons and hence the evolution of the photon spectrum can be
decoupled from the evolution of the number density of free electrons during
standard recombination and subsequent epochs of reionization. But the spectrum
does evolve and, assuming the deviations from a pure black-body are small, as
they will be for cases under consideration here, the evolution of the spectrum
can be studied in terms of the spectral distortions, known as the  Compton
$y$-parameter or $y$-distortion \cite{Zeld:1969,Chan:1975,Weymann:66,Burigana:95},
the free-free distortion \cite{Danese:77,Stebbins:86} and the
$\mu$-distortion \cite{Zeld:1969,Chan:1975,Weymann:65}. The COBE
satellite had on board an instrument, known as 
FIRAS, which measured the spectrum of the CMB to high degree of accuracy,
placing apparently stringent upper bounds
on these distortions. Subsequently, it was 
shown these upper bounds placed constraints on energy input at epochs before
the standard time of recombination, but as we will show they place only very
weak limits on the type of energy input that we have discussed in the previous 
section.   

The source of energy input we have introduced will only lead to free-free and
$y$-distortions, with the $\mu$-distortion zero, if there is no significant
energy input before the time of standard recombination, since the rate of
Compton scattering is much slower than the expansion at these
redshifts \cite{Hu:1993}. Similarly, we can ignore double Compton scattering relative to
single Compton scattering after recombination \cite{Hu:1993}. Hence, we need only consider the
$y$-distortion and the free-free distortion. Compton scattering conserves the
total number of photons and hence the only way to increase the energy of the
radiation due to the interaction with the hot plasma is by redistributing the
photons from lower frequencies to higher frequencies. This is measured using
the $y$-distortion. The thermal bremsstrahlung process on the other hand
produces photons in the low frequency region, creating the free-free
distortion. 
If the temperature is measured in the Rayleigh-Jeans part of the spectrum, it
deviates due to the increased/decreased number of photons in this part of the
spectrum by \cite{Zeld:1969,Bart:1991}
\beq
\begin{array}{lcl}
\displaystyle\frac{T_{\rm RJ}(x) - T_0}{T_0} &=& -2y \,,\\ & & \\
\displaystyle\frac{T_{\rm RJ}(x) - T_0}{T_0} &\simeq&
\displaystyle\frac{\yff}{x^2}\,, 
\end{array}
\eeq
where $T_{\rm RJ}(x)$ is the Rayleigh-Jeans temperature and $T_0$ is the
measured temperature of the Planckian distribution. 

\begin{figure}[h]
\setlength{\unitlength}{1cm}
\centerline{\hbox{\psfig{figure=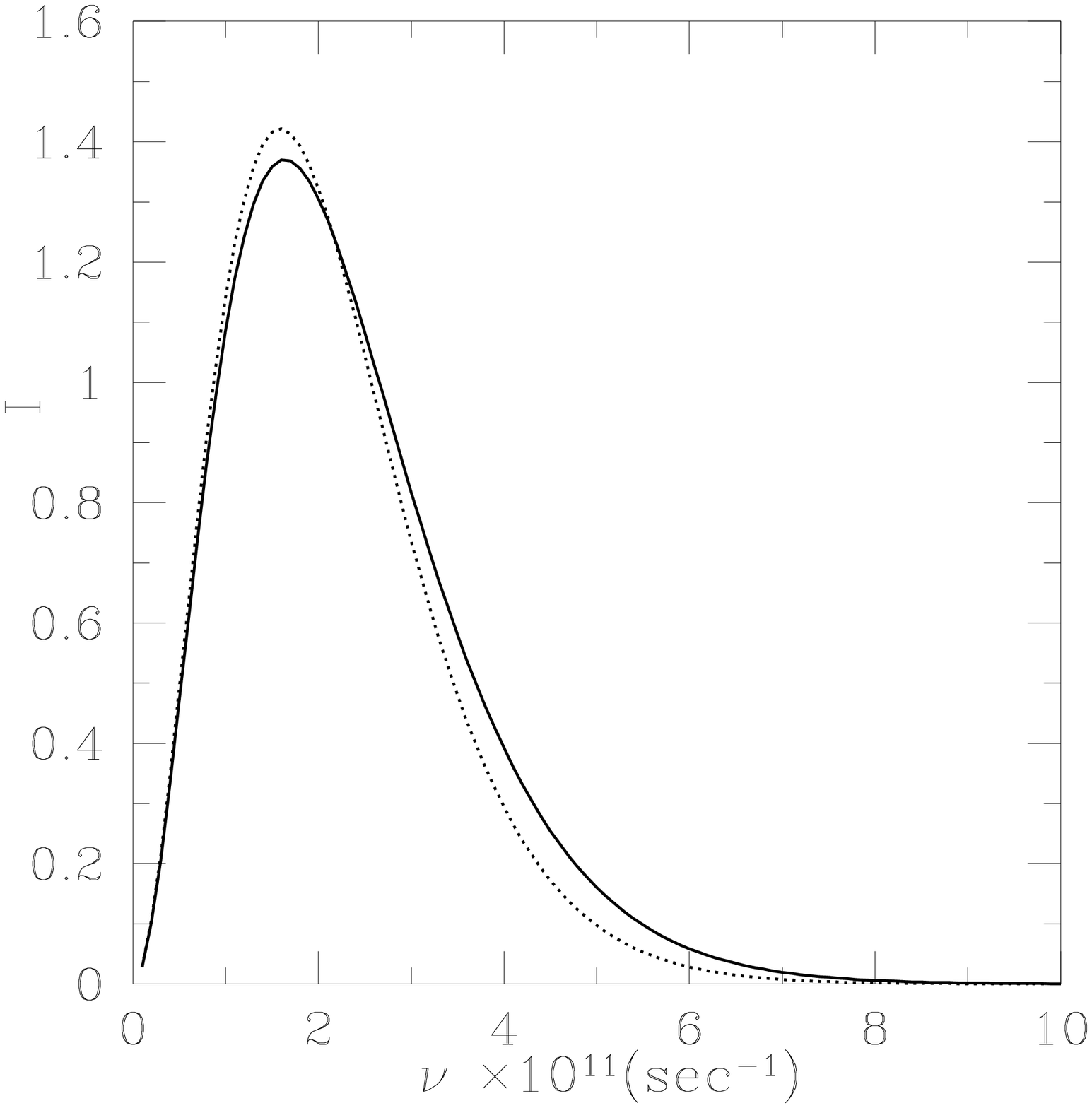,height=5cm,width=5cm}
\psfig{figure=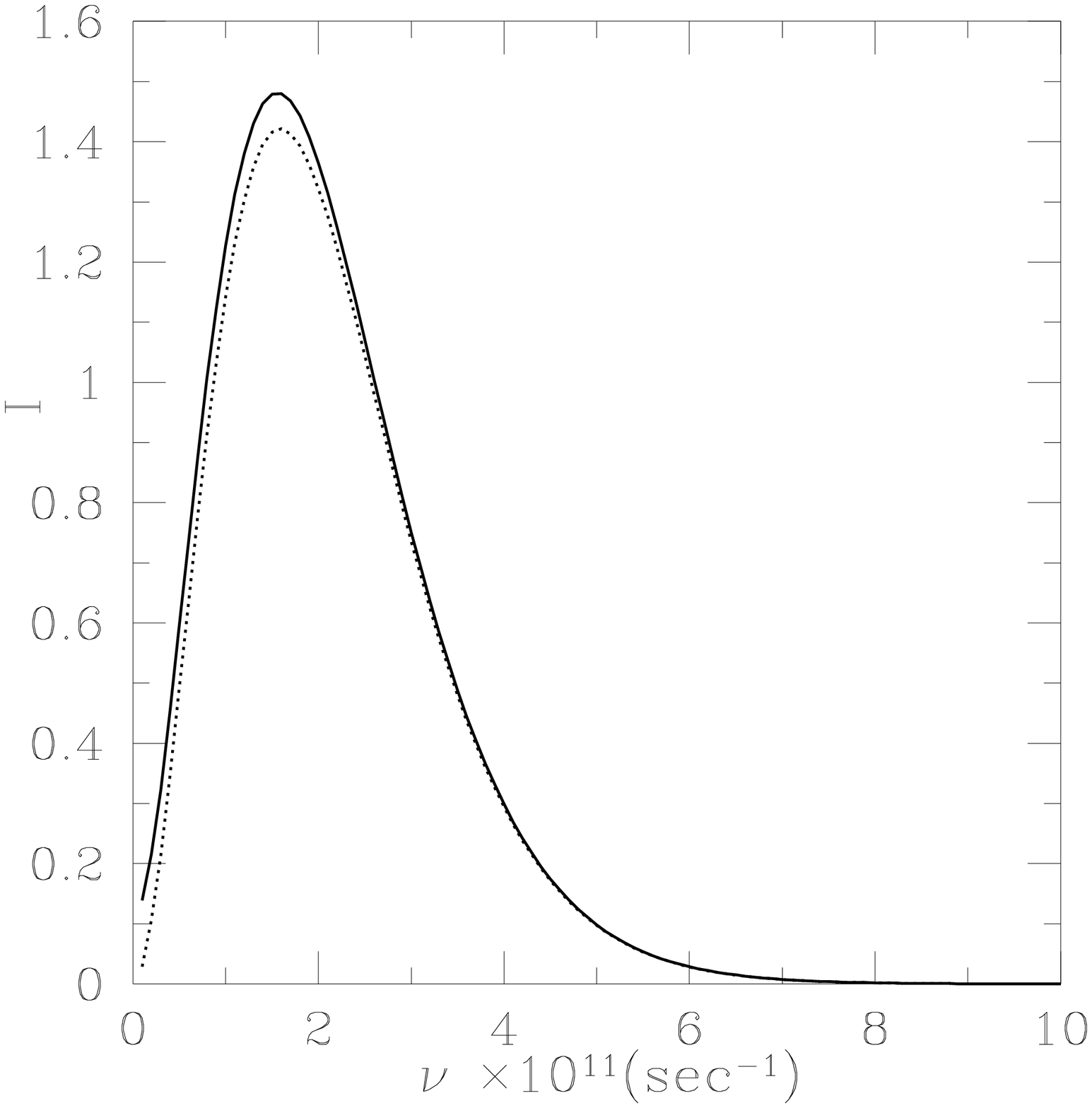,height=5cm,width=5cm}\psfig{figure=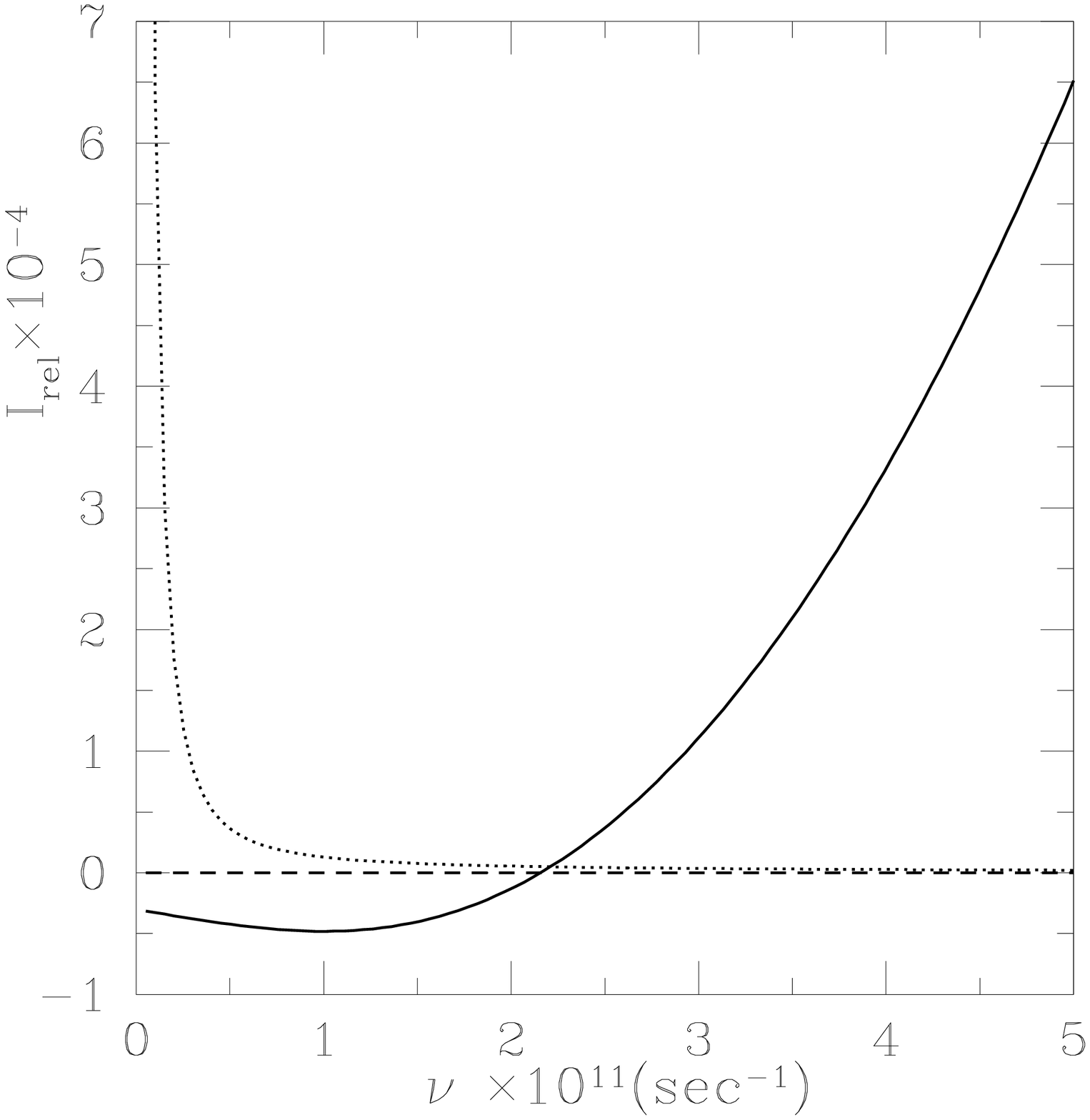,height=5cm,width=5cm}}}

\caption{\label{spectrum} On the left a y-distorted (solid line) and a
Planckian (dotted line) spectrum are
plotted, where it can be seen that the photons are redistributed from lower
frequencies, to higher frequencies. In middle a free-free distorted spectrum
(solid line) is
plotted where the increased number of low-frequency photons is established. On
the right the distortions with the value of the upper limits from the COBE FIRAS experiment are plotted (solid
line y-distortion, dotted line free-free distortion) relative to  
the related Planckian spectrum ($(n-n_{\mbox{\tiny Pl}})/n_{\mbox{\tiny Pl}})$).}
\end{figure}

The COBE FIRAS limits \cite{Smoot:1996} on these two parameters are $|y|
< 1.5 \times 10^{-5}\; (95\%{\rm CL})$ and $\left| \yff \right| < 1.9
\times 10^{-5}\; (95\% {\rm CL})$. In fig. (\ref{spectrum},left) we have
plotted the brightness $I$ ($I\propto x^3 n(x)$, with $x \propto \nu/T$) of a spectrum with $y = 1.5 \times 10^{-2}$. The redistribution of low
frequency 
photons to higher frequencies, due to the interaction with the hot plasma, is
evident from this graph. In fig. (\ref{spectrum},middle) the brightness of a 
spectrum with a 
distortion due to the release of bremsstrahlung photons from the plasma is
shown, with $\yff = 0.1$. Note that the distorted graph in
fig. (\ref{spectrum},middle) is only valid in the Rayleigh-Jeans
limit. However, one can see the increasing brightness due to extra photons in the low frequency 
region. Fig. (\ref{spectrum},right) shows the relative distortion of the
brightness, $I_{\rm rel}=[I(x)-I^{\rm Pl}(x)]/I^{\rm Pl}(x)$ where $I^{Pl}(x)$ is the brightness of the Planckian spectrum. The values of the distortions in this plot are taken from the upper limits from the COBE FIRAS experiment. 

The   
creation of spectral distortions is dictated by the photon Boltzmann equation (\ref{boltphoton}) sourced by the Kompaneets term for Compton scattering \cite{Komp:1957,Zeld:1969,Sun:1969,Chan:1975}
\beq
\left.\frac{\partial n}{\partial t} \right|_{\rm cs} =
\nef \sigma_{\mbox{\tiny T}} \frac{T_\gamma}{\me}  \frac{1}{x^2} 
\frac{\pd}{\pd x} \left(x^4 \left[\frac{\Te}{T_\gamma} \frac{\pd n}{\pd x} 
+n(1+n) \right]\right)\,,
\label{komp}
\eeq
and free-free bremsstrahlung via \cite{Komp:1957,Chan:1975}
\beq
\left. \frac{\partial n}{\partial t} \right|_{\rm br} = \kappa \frac{e^{-x_p}}
{x^3} \left[1-n\left(e^{x_p}-1\right)\right]\,,
\label{freefree}
\eeq
where $x=2\pi\nu/T_\gamma$ and $x_p=2\pi\nu/T_{\rm e}$ are dimensionless frequencies
relative to $T_\gamma$  the unperturbed temperature of the CMB and the electron
temperature $T_{\rm e}$. The coefficient $\kappa$ is given by, 
\beq
\kappa={32\pi^3e^6\nef\nb g_{\mbox{\tiny{br}}}(x_p) \over 3\me
T_\gamma^3\sqrt{6\pi\me\Te}}\,, 
\eeq 
for a plasma containing both hydrogen and helium \cite{Hu:1993} and
$g_{\mbox{\tiny br}}(x_p)$ is a Gaunt factor, accounting for quantum 
corrections to the free-free radiation process. One can either use a frequency
averaged Gaunt factor $g_{\mbox{\tiny br}}(x_p)=\overline{g}_{\mbox{\tiny
br}}\approx 1.2$ which gives an error less than $20\%$ \cite{Ryb:1979}, or a
more precise definition  $g_{\mbox{\tiny br}}(x_p) =
\sqrt{3}\ln\left(2.25/x_p\right)/ \pi$ for $x_p 
\leq 0.37$ and $g_{\mbox{\tiny br}}(x_p) = 1$ for $x_p \geq 0.37$
\cite{Hu:1993}.  

To account for the distortions in the black-body spectrum, we solve the
Boltzmann equation in terms of the unperturbed spectrum $n_0(x)$ and the
spectral distortions $y$ or $\yff$, that is, $n(x)=n_0(x)+yf(x)$,
where \cite{Zeld:1969}  
\beq
f(x)={x e^{x}\over (e^x-1)^2}\left( {x\over\tanh(x/2)}-4\right)\,.
\eeq
If we do this in an $\Omega=1$ universe then the $y$ distortion is given by
\beq 
y\equiv \int\limits_{t_0}^{t_{\rm h}} \sigma_{\mbox{\tiny T}} \nef
\displaystyle\left(\frac{\Te-T_\gamma}{\me}\right) dt
\approx 1.16 \times 10^{-11} \omb h\left(1-\yh\right)\int\limits_{0}^{z_{\rm h}}
\left({\Te - T_\gamma\over 1{\rm K}}\right)\fr(z)(1+z)^{1/2} dz\,,
\eeq
and the free-free distortion is given by \cite{Bart:1991}
\beq
\yff \equiv \int\limits_{t_0}^{t_{\rm h}} \kappa \displaystyle\frac{\Te
-T_\gamma}{\Te} dt 
\approx 2.33 \times 10^{-6} \overline{g}_{\mbox{\tiny br}} \omb^2
h^3\left(1-\yh\right) \int\limits_0^{z_{\rm h}}
\left(1-\frac{T_0}{\Te}(1+z)\right) 
\left(\frac{\Te}{1 {\rm K}}\right)^{-1/2} \fr(z)(1+z)^{1/2} dz\,,
\eeq
where $t_{\rm h}$ and $z_{\rm h}$ are the time and redshift when the energy
input becomes significant. 

One can estimate an upper bound on the distortions due to our energy input,
assuming that we achieve total ionization, that is, $\fr=1$ and a constant
temperature, $\Te$ over a finite range, say $z=z_{\rm h}$ to $z=0$, with
$z_{\rm h}>>1$. We find that the $y$-distortion is given by \cite{Tegmark:94}
\beq
y\approx 2.3\times 10^{-5}\left({\omb\over 0.05}\right)\left({h\over
0.5}\right) 
\left({1-\yh\over 0.76}\right)\left({\Te\over 5000{\rm K}}\right)\left({z_{\rm
h}\over 1000}\right)^{3/2}
\,,
\label{apydist}
\eeq
and the free-free distortion is \cite{Bart:1991}
\beq 
\yff\approx 6.3\times 10^{-12}\left({\overline{g}_{\mbox{\tiny br}}\over 1.2}
\right)\left({\omb\over 0.05}\right)^{2}\left({h\over 0.5}\right)^{3}\left(
{1-\yh\over 0.76}\right)\left({\Te\over 5000{\rm K}}\right)^{-1/2}
\left({z_{\rm h}\over 1000}\right)^{3/2}\,.
\label{apyff}
\eeq
Since our heating model does not result in larger values than $\Te=5000\, {\rm K}$, it is easy to see that it would be difficult to achieve much more than 
$y\approx 10^{-5}$ and $\yff\approx 10^{-7}$ using standard cosmological
parameters and therefore, the limits on spectral distortions from FIRAS do not
constrain our model to any great degree. In fact, the actual spectral
distortions are likely to be smaller than the upper bounds (\ref{apydist}) and
(\ref{apyff}) since the ionization fraction and electron temperature are at
their maxima only for a small portion of the range. We have calculated these
distortions for the four models which we have been considering. They are
$y\approx 3.2\times10^{-6}$ and $\yff\approx 4.3\times 10^{-8}$ for {\it
delayed} last scattering, $y\approx 3.0\times 10^{-7}$ and $\yff\approx
5.2\times 10^{-9}$ for {\it double} last scattering, $y\approx 1.7\times
10^{-6}$ and $\yff\approx 2.6\times 10^{-8}$ for {\it shifted} last scattering,
and finally $y\approx 9.8\times 10^{-7}$ and $\yff\approx 1.6\times 10^{-8}$
for {\it total ionization}. 
For comparable heating temperatures these are also the
approximate distortions given in \cite{Tegmark:94,Bart:1991}, however
\cite{Bart:1991} does not consider a particular model for the heat
input. The heating temperatures in the model used in
\cite{Tegmark:94} are higher than the ones discussed in this paper,
resulting in much stronger limitations when reionization can occur.
Similar results are obtained in \cite{Zeld:1969,Entel:85} under the 
assumption of stationary heating.
Future measurements of the cosmic microwave background
spectrum, like the experiments on the satellite missions
MAP (Microwave Anisotropy Probe) \cite{map}, Planck Surveyor
\cite{planck,Bersanelli:96} and DIMES (Diffuse Microwave Emission Survey)
\cite{dimes,Kogut:96}, are at the level of $0.1\%$ accuracy and
can provide more stringent bounds on the distortions and could rule
out some of the models discussed here. However this depends how
accurately one can remove the galactic dust contamination from the
data \cite{Stebbins:96}.

\section{Angular power spectra of temperature anisotropies and polarization}

\subsection{Simple analytic arguments}\label{analytic}

To predict CMB anisotropies one has to solve the evolution
equation for each species of particles either numerically
\cite{Vittorio:84,Bond:84,cmbfast,Peebles:70,Wilson:81} or analytically \cite{Doroshkevich:78,Doroshkevich:88,Naselskii:93,Atrio:94,Jorgensen:95}.
One can understand the qualitative structure of  temperature anisotropies and
polarization using a formalism first applied to passive adiabatic models
\cite{Hu:95a,HSb,Zaldarriaga:95,Hu:97a} and later adapted to incorporate active source models
\cite{MACFa,MACFb,Battye:96}.  For the purpose of this section we shall ignore
the vector and tensor contributions to the anisotropies, and concentrate on the
scalar component since it is this which is most affected by the modifications
to  the thermal history discussed earlier. In this case, the angular power
spectra for the temperature anisotropies, $C^T_l$, and the
polarization\footnote{Note that since we have ignored the vector and tensor
components of the source, 
there is no magnetic component of the polarization \cite{Seljak:97}.},
$C^E_l$  are given by \cite{Hu:95a,Zaldarriaga:95}
\beq
C_l^T=\displaystyle{2\over\pi}\int_0^{\infty}k^2dk\langle\Delta_l(k,\eta_0)
\Delta_l^*(k,\eta_0)\rangle\,,\quad C_l^E=\displaystyle{2\over\pi}\int_0
^{\infty}k^2dk\langle E_l(k,\eta_0)E^*_l(k,\eta_0)\rangle\,, 
\eeq
where $\Delta(k,\eta_0,\mu)$ and $E(k,\eta_0,\mu)$ are the temperature and
polarization distribution functions expanded in terms of spherical harmonics
of the angular variable $\mu=\cos\theta$, with coefficients
$\Delta_l(k,\eta_0)$ and $E_l(k,\eta_0)$. The basic procedure involves writing
these multipoles in terms of an integral over the lower multipoles and the
gravitational potentials along the line of sight.  

The line of sight integral for the temperature anisotropies is then \cite{Hu:95a}
\beq
\Delta_l(k,\eta_0)=\displaystyle\int_0^{\eta_0} d\eta\left(S_T^0(k,\eta)
j_l\left[k\left(\eta_0-\eta\right)\right]+S^1_T(k,\eta)j_l^{10}\left[
k\left(\eta_0-\eta\right)\right]+S^2_T(k,\eta)j_l^{20}\left[
k\left(\eta_0-\eta\right)\right]\right)\,,
\label{tempdist}
\eeq
where $j_l^{10}(x)=j_l^{\prime}(x)$,
$j_l^{20}(x)=(3j^{\prime\prime}(x)+j_l(x))/2$, 
\beq
S_T^0(k,\eta)=g(\eta)\left(\Delta_0+\Psi\right)+h(\eta)\left(\dot\Psi-\dot\Phi
\right)\,,\quad S_T^1(k,\eta)=g(\eta)V_{\rm B}\,,\quad
S_T^2(k,\eta)=g(\eta)P\,, 
\eeq
$\Psi,\Phi$ are the gravitational potentials characterizing the effects of the
sources, $V_{\rm B}$ is the baryon velocity and $P=(\Delta_2-E_2)/2$. The
equivalent expression for the electric component of the polarization
distribution is \cite{Hu:97a,Hu:98b}
\beq
E_l(k,\eta_0)=-\sqrt{6}\displaystyle\int_0^{\eta_0}d\eta
g(\eta)P(k,\eta)\epsilon_l\left[k\left(\eta_0-\eta\right)\right]\,, 
\label{polardist}
\eeq
where 
\beq
\epsilon_l(x)=\sqrt{{3\over 8}{(l+2)!\over (l-2)!}}{j_l(x)\over x^2}\,.
\eeq
For the standard thermal history the last scattering visibility function can
be approximated for pedagogical purposes by a
delta function at $\eta_*$ and the cumulative visibility function by a step
function at the same point, that is, $g(\eta)=\delta(\eta-\eta_*)$ and
$h(\eta)=\Theta(\eta-\eta_*)$. Hence, (\ref{tempdist}) and
(\ref{polardist}) can be approximated by \cite{Hu:95a,Hu:97a}
\bea
\Delta_l(k,\eta_0)=&\left(\Delta_0+\Psi\right)(k,\eta_*)j_l\left[k
\left(\eta_0-\eta_*\right)\right]+V_{\rm B}(k,\eta_*)j_l^{10}\left[k\left(
\eta_0-\eta_*\right)\right]+P(k,\eta_*)j_l^{20}\left[k\left(\eta_0-\eta_*
\right)\right] \nonumber \\
&+\displaystyle\int_{\eta_*}^{\eta_0}d\eta\left(
\dot\Psi-\dot\Phi\right)j_l\left[k\left(\eta_0-\eta\right)\right]\,,
\label{approxsol}
\eea
and  
\beq
E_l(k,\eta_0)=-\sqrt{6} P(\eta_*)\epsilon_l\left[k\left(\eta_0-\eta_*\right)
\right]\,.
\eeq
Therefore, one can investigate the qualitative nature of the anisotropies
and polarization by estimating $\Delta_0+\Psi$, $V_{\rm B}$ and $P$ around the
time of last scattering\footnote{In fact, we will calculate the
$\Delta_0+\Phi$ around the time of last scattering and assume that $\Psi-\Phi$ is small.} and $\dot\Psi-\dot\Phi$ along the line of sight. 

We shall primarily be interested in the contribution from acoustic
oscillations, since it is they which are most sensitive to the
ionization history. Using the tight coupling approximation, that is,
an expansion in powers of $1/\dot\tau$, one can deduce that \cite{Hu:95a}  
\bea
\widehat\Delta_0(\eta_*)=&\Delta_0(\eta_*)+\Phi(\eta_*)=e^{-k^2/k^2_{\rm
s}(\eta_*,0)} 
\widehat\Delta_0(0)\cos\left(\displaystyle
{k\eta_*\over\sqrt{3}}\right)+\displaystyle{\sqrt{3}\over
k}\dot{\widehat\Delta}_0(0)\sin\left({k\eta_*\over\sqrt{3}}\right)
\nonumber \\ 
&+\displaystyle{\sqrt{3}\over k}\int_0^{\eta_*}d\eta^{\prime}e^{-k^2/k^2_{\rm
s}(\eta_*,\eta^{\prime})} 
\sin\left({k\over\sqrt{3}}(\eta_*-\eta^{\prime})\right)F(\eta^{\prime})\,, 
\label{tight}
\eea
where we have ignored the effects of baryons, $F(\eta)=k^2(\Phi-\Psi)/3$ is
the structure function of the source,  
\beq
k_{\rm s}^{-2}(\eta_2,\eta_1) =\displaystyle{4\over
27}\int_{\eta_1}^{\eta_2}{d\eta\over\dot\tau(\eta)}\,, 
\label{silk}
\eeq
is the damping length due to photon diffusion
\cite{Kaiser:84,Hu:97a,Silk:68} and ${\widehat\Delta}_0(\eta)\equiv\Delta_0(\eta)+\Phi(\eta)$. This serves as an approximation for the intrinsic anisotropy created around the time of recombination. Furthermore, the other
important quantities $V_{\rm B}$ and $P$ are related to $\widehat\Delta_0$ by
$P\sim V_B\sim\Delta_1\sim\dot{\widehat{\Delta}}_0$ around this time.  

With a few subtleties, the three parts to the tight coupling solution
(\ref{tight}) correspond to the passive adiabatic $(\dot{\widehat\Delta}_0(0)=0, F(\eta)=0)$, passive isocurvature $({\widehat\Delta}_0(0)=0, F(\eta)=0)$ and
active sources $({\widehat\Delta}_0(0)=0, \dot{\widehat\Delta}_0(0)=0)$. When all the relevant
effects are taken into account the contribution from $V_B$ is suppressed
relative to that from  $\Delta_0+\Psi$, and that from $P$ is even
further suppressed. Therefore, there are in general two oscillatory components to the temperature anisotropy which are out of phase with each other, one which gives rise to peaks, $\Delta_0+\Psi$, and the other which fills in between them creating troughs, $V_B$. These arguments lead to the much discussed {\it acoustic peaks} in the standard adiabatic scenario, which has the first peak at around $l=200$ corresponding to the size of the sound horizon at time of recombination. Isocurvature and active source models also have similar peak structures\footnote{We are only considering coherent active source models at this stage.}, albeit with the main peak at slightly larger $l$\cite{Hu:96c}.
The effective source of polarization $\Delta_1$ is out of phase with that for the peaks in the
anisotropy spectrum and there is no contribution to the polarization from their
source, $\Delta_0+\Psi$. This leads to a set of tight peaks which are out of
phase with those for anisotropy. The amplitude of the polarization is
generally much lower than the anisotropy since no net polarization is created
during the tight coupling epoch \cite{Kaiser:84,Bond:84,cmbfast,Hu:98b,Ng:95}.   
We wish to modify this simple qualitative treatment of the structure
of the anisotropies to the case where we have a more complicated
ionization history. In order to do this,  we replace the last
scattering visibility function by \cite{Zald:96,Hu:97b}
\beq
g(\eta)=e^{-\tau(\eta_{\rm r})}\delta(\eta-\eta_*)+\left(1-e^{-\tau(\eta_{\rm r})}\right)\delta(\eta-\eta_{\rm r})\,,
\label{twolastscat}
\eeq
where $\eta_{\rm r}$ is the time of reionization. Using this one can deduce   
that the cumulative visibility function will be given by $h(\eta)=0$ for $\eta<\eta_*$, $h(\eta)=e^{-\tau(\eta_{\rm r})}$ for $\eta_*<\eta<\eta_{\rm r}$ and $h(\eta)=1$ for $\eta>\eta_{\rm r}$. This is similar to the case of {\it double} last scattering discussed in the previous section which is the most general case. However, we shall discuss how this approach can be modified to understand the effects of the {\it shifted} and {\it delayed} last scattering scenarios. Since the visibility functions are not well represented in this way for the case of the {\it total ionization} scenario, we shall only comment briefly on its effects in this section.

Using (\ref{twolastscat}) we can deduce that \cite{Zald:96}
\bea
\Delta_l(k,\eta_0) &=& e^{-\tau(\eta_{\rm
r})}\bigg{[}\left(\Delta_0+\Psi\right)(k,\eta_*)j_l\left[k(\eta_0-\eta_*)
\right]+ V_{\rm
B}(k,\eta_*)j_l^{10}\left[k(\eta_0-\eta_*)\right]+P(k,\eta_*)j_l^{20}\left[k(\eta_0-\eta_*)\right]\bigg{]} 
\nonumber \\
&+&\left(1-e^{-\tau(\eta_{\rm
r})}\right)\bigg{[}\left(\Delta_0+\Psi\right)(k,\eta_{\rm
r})j_l\left[k(\eta_0-\eta_{\rm r})\right]+V_{\rm B}(k,\eta_{\rm
r})j_l^{10}\left[k(\eta_0-\eta_{\rm r})\right]+P(k,\eta_{\rm r})j_l^{20}\left[k(\eta_0-\eta_{\rm r})\right]\bigg{]} \nonumber \\
&+&e^{-\tau(\eta_{\rm r})} \int_{\eta_*}^{\eta_{\rm r}}
d\eta \left(\dot\Psi-\dot\Phi\right)j_l\left[k(\eta_0-\eta)\right] +
\int_{\eta_{\rm
r}}^{\eta_0}d\eta\left(\dot\Psi-\dot\Phi\right)j_l\left[k(\eta_0-\eta)\right]\,
.\label{2LSS}
\eea
The first term in square brackets is due to photons which were last scattered at the time of recombination. Only a fraction $e^{-\tau(\eta_{\rm r})}$ of the total number of photons observed today will have not scattered since that epoch and the remainder, a fraction $1-e^{-\tau(\eta_{\rm r})}$, was scattered at the time of reionization. The effect of these photons on the temperature anisotropy is given by the second term. The final two terms are a similarly modified version of the 
integrated Sachs-Wolfe (ISW) effect.

We have already estimated the effects of $\Delta_0+\Psi$ and $V_B$ at the time of recombination, and these are just damped by the factor $e^{-\tau(\eta_r)}$ in these scenarios. However, we must now also estimate their effects at the time of reionization. We assume that the photons free stream from the time of recombination until just before reionization at time $\eta_{\rm r}-\epsilon$, where $\epsilon$ is usually small when compared to $\eta_{\rm r}$. This yields
\bea
\Delta_0(\eta_{\rm r}-\epsilon,k)=&\left(\Delta_0+\Psi\right)(k,\eta_*)j_0\left[k
\left(\eta_{\rm r}-\epsilon-\eta_*\right)\right] +\displaystyle\int_{\eta_*}^{\eta_{\rm r}-\epsilon}d\eta\left(
\dot\Psi-\dot\Phi\right)j_0\left[k\left(\eta_{\rm r}-\epsilon-\eta\right)\right]\,.
\label{approxsoleps}
\eea
If we now assume that the ISW component to this is negligible, as it
will be in most applications, then this simple multiplication can be
used as an initial condition for the epoch when the photons re-enter
the phase where the
photons couple again to the electrons. Using (\ref{tight}), one can deduce
that \cite{Zald:96}  
\bea
\widehat\Delta_0(\eta_{\rm r})=&\Delta_0(\eta_{\rm r})+\Phi(\eta_{\rm r})=e^{-k^2/k^2_{\rm
s}(\eta_{\rm r},\eta_{\rm r}-\epsilon)} 
\bigg{[}\widehat\Delta_0(\eta_{\rm r}-\epsilon)\cos\left(\displaystyle
{k\epsilon\over\sqrt{3}}\right)+\displaystyle{\sqrt{3}\over
k}\dot{\widehat\Delta}_0(\eta_{\rm r}-\epsilon)\sin\left({k\epsilon\over\sqrt{3}}\right)\bigg{]}
\nonumber \\ 
&+\displaystyle{\sqrt{3}\over k}\int_{\eta_{\rm r}-\epsilon}^{\eta_{\rm r}}d\eta^{\prime}e^{-k^2/k^2_{\rm
s}(\eta_{\rm r},\eta^{\prime})} 
\sin\left({k\over\sqrt{3}}(\eta_{\rm
r}-\eta^{\prime})\right)F(\eta^{\prime})\, .
\label{tight2LSS}
\eea
Computing $V_B$  and $P$ around the time of reionization is more tricky since
there is pre-existing anisotropy. If this is large then it can modify the
relation between $V_B$, $P$ and $\Delta_1$. We do not believe that this will
have a substantial effect on the temperature, but we shall return to this point
when we discuss polarization below. 

We are now in a position to discuss the effects of this {\it double} last
scattering scenario on the CMB anisotropies. The fact that there are two last
scattering surfaces will lead to two sets of peaks with relative amplitudes
$e^{-\tau(\eta_{\rm r})}$ and $1-e^{-\tau(\eta_{\rm r})}$, whose scales will be
set by the sound horizons at the time of recombination $(k_*\sim \eta_*^{-1}$)
and the time of reionization. By substituting the free-streamed tight coupling
solution into (\ref{tight2LSS}),  one can deduce that this scale is $k_r\sim
(\eta_*+\epsilon)^{-1}$, which implies that the size of the sound horizon is
proportional to the time for which acoustic oscillations take place. Remember
that $\epsilon$ is small relative to $\eta_{\rm r}$, but so long as $\eta_{\rm
r}\gg\eta_*$, then it is possible for $\epsilon\sim{\cal O}(\eta_*)$,
creating peaks which are
based around very different scales. These scales can be projected into
$l$-space \cite{Hu:95a} using the fact that the spherical Bessel function $j_l(x)$ is peaked
around $x=l$, which in the case of the anisotropy created at the time of last
scattering is $l\approx k_*(\eta_0-\eta_*)$ and for that created at
reionization is $l\approx k_r(\eta_0-\eta_r)$. 

This qualitative approach can be modified to also explain the nature of the
structure of the anisotropy in the {\it shifted} last scattering scenario. In
this case, the optical depth of the reionization is very large, $\tau(\eta_{\rm
r})\gg1$, which implies that one can ignore the first term in (\ref{2LSS}), and
hence the anisotropy is given by  
\bea
\Delta_l(k,\eta_0) = & \left(\Delta_0+\Psi\right)(\eta_{\rm
r})j_l\left[k(\eta_0-\eta_{\rm r})\right]+V_{\rm B}(\eta_{\rm
r})j_l^{10}\left[k(\eta_0-\eta_{\rm r})\right]+P(\eta_{\rm
r})j_l^{20}\left[k(\eta_0-\eta_{\rm r})\right] \nonumber \\
&+\displaystyle{\int_{\eta_{\rm
r}}^{\eta_0}d\eta\left(\dot\Psi-\dot\Phi\right)j_l\left[k(\eta_0-\eta)\right]}
\,. 
\label{singleLSS}
\eea
Furthermore, one can assume that the tight coupling regime effectively never
ended at the time of recombination, which requires us to make the approximation
$\eta_{\rm r}\approx \epsilon$ (and $\eta_*=0$). This is clearly not totally true, since recombination of the protons and electrons did take place, but the only thing critical for estimating the anisotropies is the visibility. In the specific case of {\it shifted} last scattering which we are considering here, the visibility of the time of recombination is almost negligible. Therefore, we can estimate 
\bea
\widehat\Delta_0(\eta_{\rm r})=&\Delta_0(\eta_{\rm r})+\Phi(\eta_{\rm r})=e^{-k^2/k^2_{\rm
s}(\eta_{\rm r},0)} 
\bigg{[}\widehat\Delta_0(0)\cos\left(\displaystyle
{k\eta_{\rm r}\over\sqrt{3}}\right)+\displaystyle{\sqrt{3}\over
k}{\dot{\widehat\Delta}}_0(0)\sin\left({k\eta_{\rm r}\over\sqrt{3}}\right)\bigg{]}
\nonumber \\ 
&+\displaystyle{\sqrt{3}\over k}\int_{0}^{\eta_{\rm r}}d\eta^{\prime}e^{-k^2/k^2_{\rm
s}(\eta_{\rm r},\eta^{\prime})} 
\sin\left({k\over\sqrt{3}}(\eta_{\rm
r}-\eta^{\prime})\right)F(\eta^{\prime})\, .
\label{tight1LSS}
\eea
We see that now that there is just one set of peaks with their scale
set by $k_{\rm r}\sim \eta_{\rm r}^{-1}$. Since $\eta_{\rm
r}\gg\eta_{*}$, this corresponds to a shift of the entire peak
structure to smaller $l$. Also the diffusion damping length $k_{\rm
s}^{-1}(\eta_{\rm r},0)$ has grown considerably and hence the effects
of Silk damping \cite{Silk:68} are prevalent on larger scales than in
the standard scenario \cite{Hu:97b}. 

We could also use this limit to understand the effects of {\it delayed}
recombination, since there is just one surface of last scattering, but it seems
more logical to treat this case as its name suggests as an increase in
$\eta_*$, with $\tau(\eta_{\rm r})=0$. This leads to a simple shift in the time of recombination, and hence the size of the sound horizon when the photons last
scatter. The observational consequence is that the whole spectrum of
anisotropies and also the Silk damping envelope are shifted to larger scales.

The effects of these modified thermal histories on the position of peaks in the polarization is very similar to their effects on the temperature anisotropy. This is since the evolution of the source of polarization inside any surface of last scattering induced by reionization is very much the same as in the standard case.  Therefore, if we make the same assumptions about the surface of last scattering, then the polarization is given by \cite{Zald:96}
\beq
E_l(k,\eta_0) = -\sqrt{6}\left\{e^{-\tau(\eta_{\rm
r})}P(\eta_*)\epsilon_l\left[k(\eta_0-\eta_*)\right] +
\left(1-e^{-\tau(\eta_{\rm r})}\right)P(\eta_{\rm
r})\epsilon_l\left[k(\eta_0-\eta_{\rm r})\right]\right\}\, .
\label{pol2LSS}
\eeq
As for the anisotropy, the function $\epsilon_l(x) \propto j_l(x)$ peaks around $l\approx x$ and the two contributions to (\ref{pol2LSS}) produce
peaks on the two scales $l\approx k_*(\eta_0-\eta_*)$ and $l\approx
k_r(\eta_0-\eta_{\rm r})$. Similar shifts in the peak position are possible in the cases of {\it shifted} and {\it delayed} last scattering. Since the polarization has only one source, as opposed to the temperature which has a number, this effect will be seen much more clearly. 

However, the amplitude of polarization on different scales is much more difficult to understand in any kind of generality. First, there are effects of Silk damping, which are likely to be very much the same as for the anisotropy. But more importantly there is the added difficulty of pre-existing polarization at the time of reionization. In the standard case, the quadrupole is negligible until very close to the time of last scattering and so its amplitude increases during that time, while oscillating out of phase with the monopole. It is the balance between this increase and the effects of Silk damping which makes the amplitude maximum at around the third or fourth peak.  If the quadrupole is non zero before the time of reionization, as is likely to be the case particularly in the {\it double} last scattering scenario, then it is possible to shift this balance to larger scales. Hence, it is possible for the maximum amplitude to be at the first peak.  
 
It is hard to obtain any analytical predictions for the total ionization 
model. This is because the last scattering visibility function is smeared out
over the whole history since recombination and can not be
approximated by a simple function with two peaks (fig.~\ref{fig:total}). Therefore the line
of sight integrals (\ref{tempdist},\ref{polardist}) can not be carried out in
a simple way. The only prediction we can make 
is, since this scenario gives a small cumulative visibility up to very late times
(see fig.~(\ref{fig:total})), last scattering visibility function), that the
resulting anisotropy power spectrum will be suppressed up to relatively large scales.

\begin{figure}[h]
\setlength{\unitlength}{1cm}
\centerline{\hbox{\psfig{figure=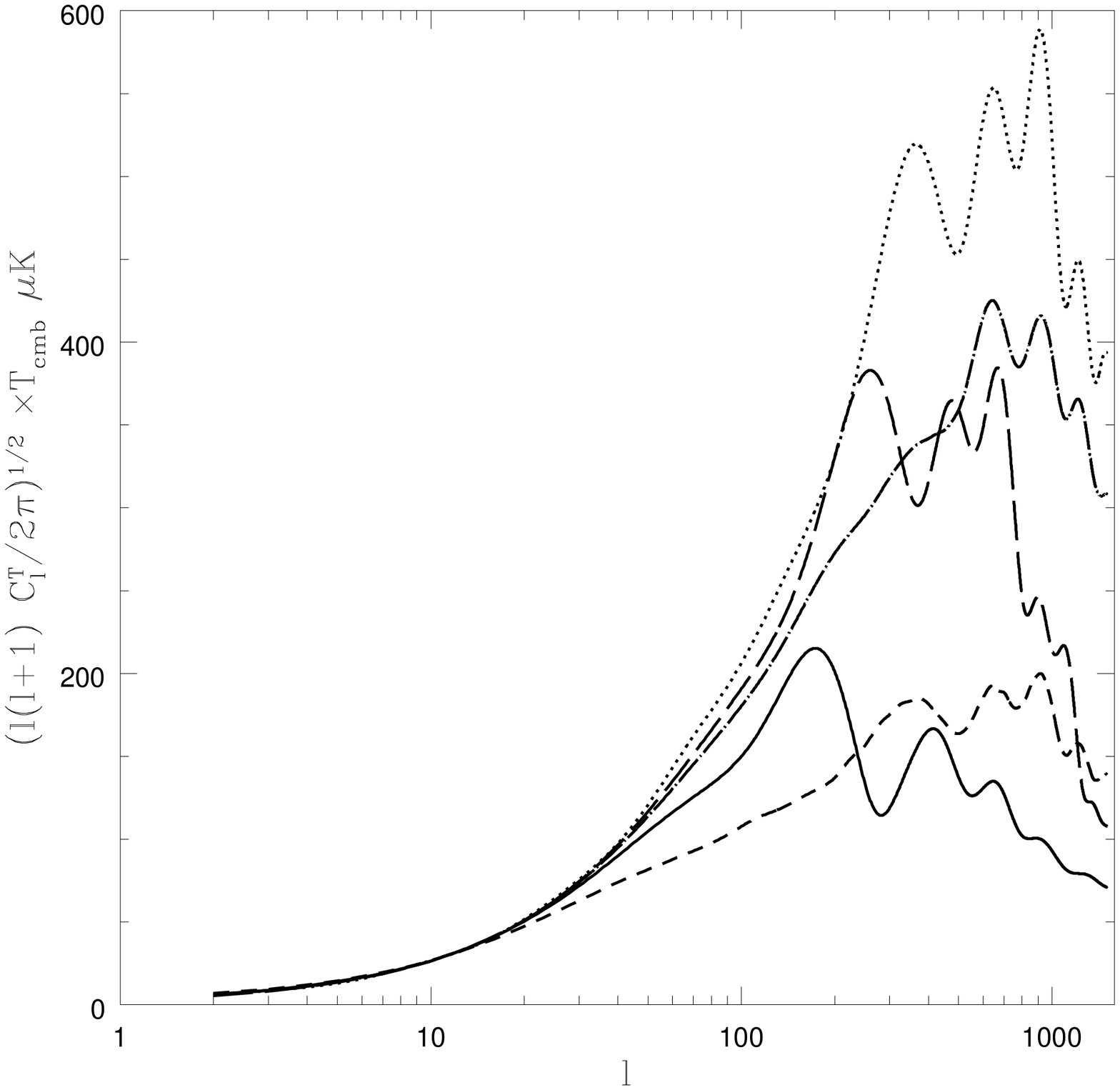,height=8cm,width=9cm}
\psfig{figure=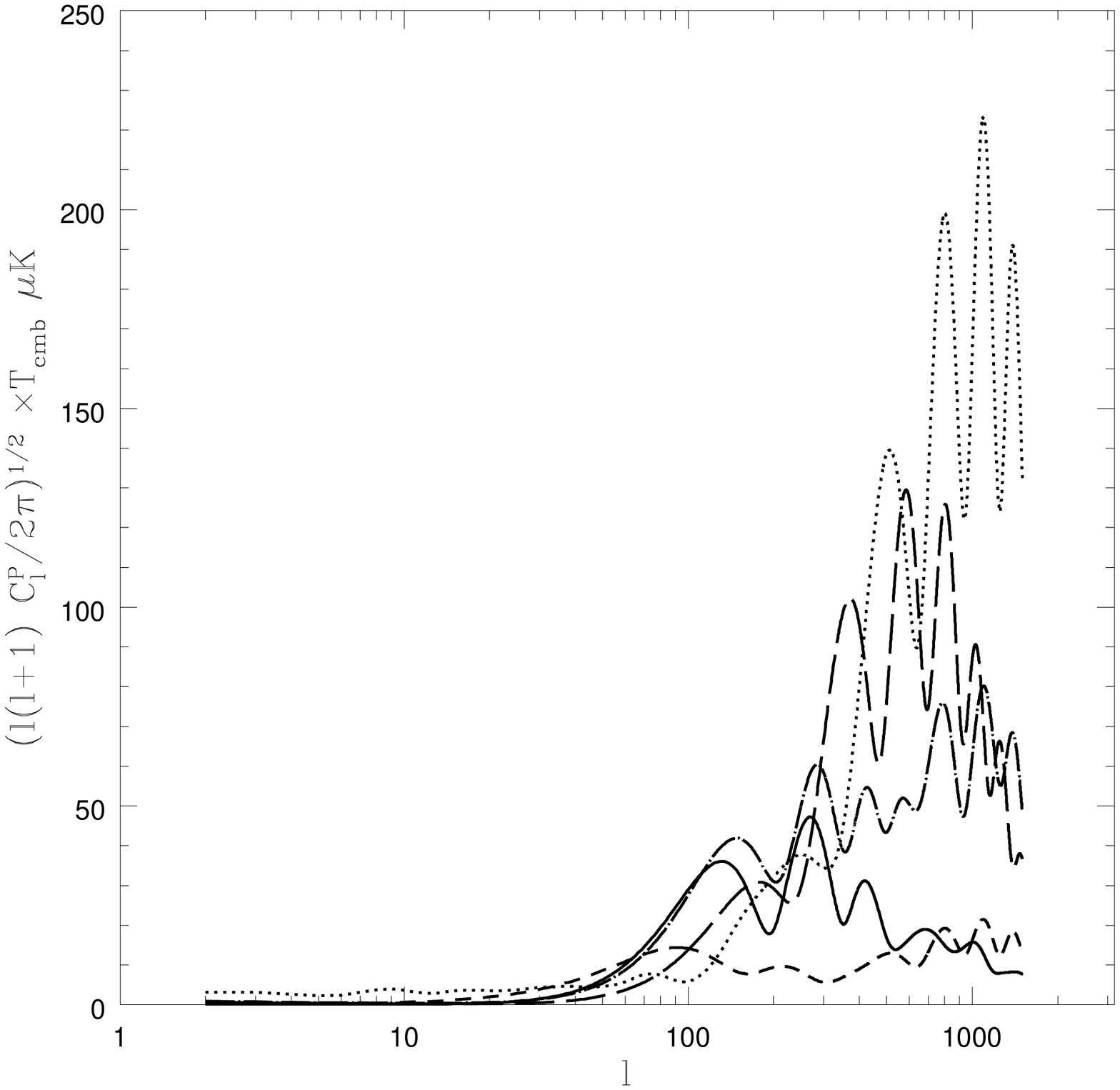,height=8cm,width=9cm}}}
\caption{\label{powerspectrum}The temperature (left) and polarization (right)
anisotropy power spectra for an isocurvature CDM
model. The dotted line refers to the standard thermal history, the long dashed
line to delayed last scattering, the solid line to shifted last scattering, the
dot-dashed line to double last scattering, and the
short dashed line to the total ionization scenario.}
\end{figure}

\subsection{Isocurvature white noise}
Our main emphasis in this paper is on active source models. However, there is
a class of passive models, known as isocurvature white-noise models, which have
a number of the features of an active model. We have integrated the linearized
Einstein-Boltzmann equations with  an integrator known as 
CMBFAST \cite{cmbfast} for an isocurvature CDM model
\cite{Linde:85,Seckel:85,Peebles:97} with the initial power
spectrum
\beq
	P_i(k) = k^0 \,,
\eeq
that is, an initial white-noise spectrum and the results for our sample of
modified thermal histories discussed in the previous section are shown in
fig.(\ref{powerspectrum}). The power spectrum is plotted up
to $l=1500$, because we do not expect interessting features on smaller
scales in this first order approach.
We should note first that, for the standard thermal history, the amplitude and
spectrum of such a model is very much in conflict with the current
observations \cite{Hu:94b}. However, as we shall discuss in section III.D, part of the
motivation for this work is to investigate how the changes in the ionization
history might create a more acceptable model.  

We see that the peak structure in the delayed and shifted last scattering
scenarios is moved to much larger scales and that amplitude is suppressed by
the damping envelope, very much in keeping with the analytic arguments of the
previous section. For the case of double last scattering the analytic treatment
suggests that there should be a second set of peaks at larger scales, but it
appears that in this particular case that this second set of peaks is
suppressed below that of the first set. The first peak which appears for the
standard case  at $l\approx 350$  also appears to be `washed' out by the
damping at the time of reionization. However, the three peaks which remain are
on exactly the same scales as those in the standard case and represent the
anisotropy created at the time of the initial epoch of recombination, suppressed
by the damping factor $e^{-\tau(\eta_{\rm r})}$. Finally the total ionization
scenario results in 
a heavy suppression of power up to very large scales ($l \approx 20$) This is
because the last scattering visibility function is smeared out over the whole
time since recombination (fig.(\ref{fig:total})). We should note that
reionization only appears to be effective on small scales and in each case the
anisotropy on scales with $l<20$ remains unchanged. 

As already discussed the peak structure of the polarization power spectrum is a
more clean test of our analytic arguments, although our understanding of the
amplitude is much less clear. In the standard case, the first peak is around
$l\approx 250$ and the damping becomes effective on scales with $l>1000$. Once
again the action of the delayed and shifted last scattering is easily
understood from our analytic arguments with just a universal shift of the
spectrum and damping scales. In the double last scattering scenario we now see
that there are two clear sets of peaks which have different scales, those at
the smaller scales being those created at the time of recombination damped by
the appropriate amount. Since the visibility function at recombination is
not completely zero, the damping phase leads to a quadrupole moment in the
temperature anisotropies. This quadrupole contribution acts as a seed for
the creation of polarization anisotropies (\ref{polardist}) during
reionization and hence the first peak in the polarization for this model is
more prominent than in the others. Finally, total ionization results in a severely damped  
polarization  spectrum, although even in this very extreme case one can just
observe the remnants of the polarization created at the standard time of
recombination. 

\subsection{Simple coherent scaling source models}
In order to study active sources we will first discuss the simple
coherent scaling source models introduced in
refs.~\cite{Turok:96,Hu:97}. Although these kind of sources are unlikely
to be realized in the early universe, they have some some illustrative
value. The scaling source is introduced as components of the 
Newtonian metric perturbations; specifically the curvature $\Phi = \Phi_{\gamma
\rm{b}} + \Phi_{\rm s}$ and the gravitational potential $\Psi =
\Psi_{\gamma{\rm b}} + \Psi_{\rm s}$, where  $\Phi_{\gamma
\rm{b}}$ and $\Psi_{\gamma{\rm b}}$ are the contributions from the photon-baryon
fluid, and $\Phi_{\rm s}$ and $\Psi_{\rm s}$ are those for the source. These sources terms can be  related to the density of the source $\rho_{\rm s}$, its velocity $v_{\rm s}$ and  anisotropic stress $\pi_{\rm s}$ by \cite{Hu:97}
\beq
\begin{array}{c}
\displaystyle
	k^2\Phi_{\rm s} = 4\pi Ga^2\left(\rho_{\rm s}+3 \frac{\dot a}{a}v_{\rm
s}/k\right) \,, \\ \\
	k^2\left(\Psi_{\rm s}+\Phi_{\rm s}\right) = -8\pi Ga^2\pi_{\rm s} \,,
\end{array}
\eeq
where the derivatives are with respect to conformal time $\eta = \int dt/a$.  The conservation equations for the seed source are \cite{Hu:97}
\beq
\begin{array}{c}
\displaystyle
	\dot{\rho}_{\rm s}+3\frac{\dot{a}}{a}\left(\rho_{\rm s}+p_{\rm
s}\right) = -kv_{\rm s}\,, \\ \\
\displaystyle
	\dot{v}_{\rm s}+4\frac{\dot{a}}{a}v_{\rm s} = kp_{\rm s} -\frac{2}{3}k
\pi_{\rm s} \,,
\end{array}
\eeq
where $p_{\rm s}$ is the pressure of the source. Clearly, there are only two independent quantities here and we can in general choose any two arbitrarily, the other two being computed by evolving these equations. One simple scaling ansatz, which we shall call the pressure source, is given by \cite{Hu:97}
\beq
\begin{array}{c}
\displaystyle
	4\pi Ga^2p_{\rm s} =
\eta^{-1/2}\frac{\sin\left(Ak\eta\right)}{\left(Ak\eta\right)}\,, \\ \\
\pi_{\rm s} = 0 \, ,
\end{array}
\eeq
where $0<A<1$ and another which we shall call the stress source has
\beq
	4\pi Ga^2 \pi_{\rm s} = \eta^{-1/2}
\frac{6}{B_2^2-B_1^2}\left[\frac{\sin\left(B_1k\eta\right)}{\left(B_1k\eta\right)}-\frac{\sin\left(B_2k\eta\right)}{\left(B_2k\eta\right)}\right]
\, ,
\eeq
where $0<(B_1,B_2)<1$, with $p_s$ given by the pressure source.
For the results presented here we have chosen to use $A=1.0$, $B_1=1.0$ and $B_2=0.5$.
\begin{figure}[!h]
\setlength{\unitlength}{1cm}
\centerline{\hbox{\psfig{figure=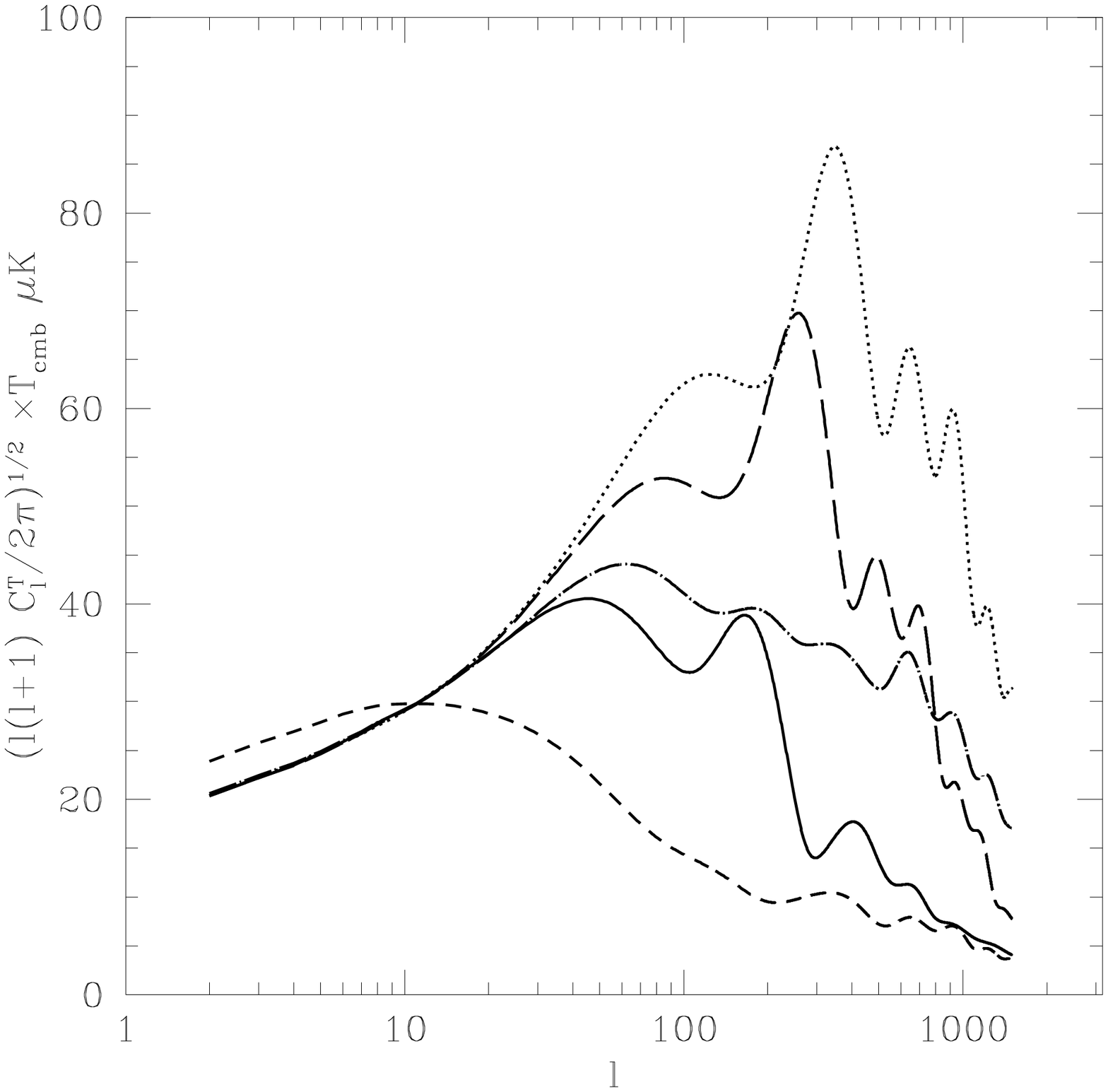,height=8cm,width=9cm}
\psfig{figure=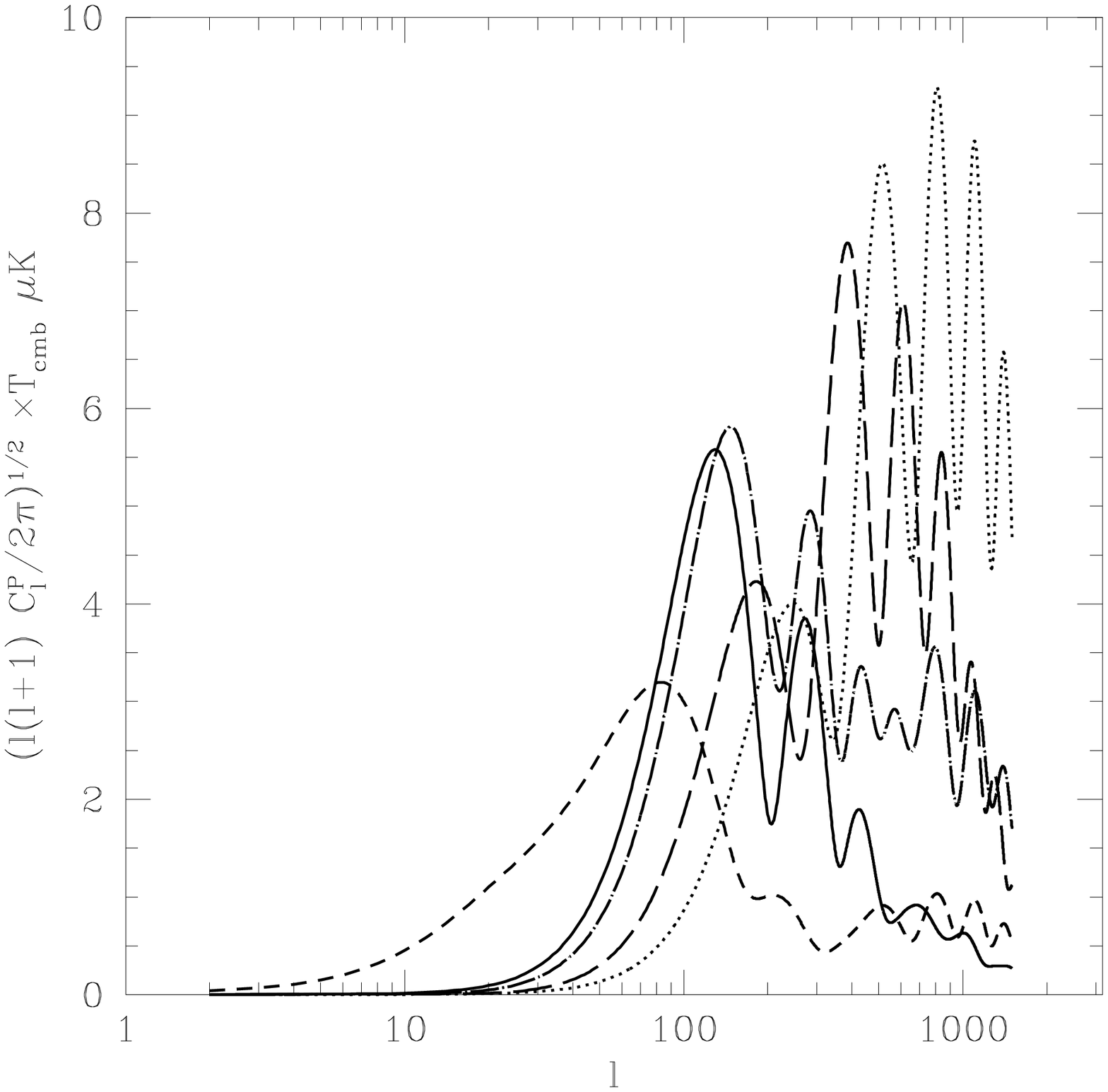,height=8cm,width=9cm}}}
\caption{\label{powerpressure}Temperature (left) and polarization (right)
anisotropy power spectra for a pressure source model with $A=1.0$. The dotted line refers to the standard thermal history, the long dashed
line to delayed last scattering, the solid line to shifted last scattering, the
dot-dashed line to double last scattering, and the
short dashed line to the total ionization scenario.}
\end{figure}
\begin{figure}[!h]
\setlength{\unitlength}{1cm}
\centerline{\hbox{\psfig{figure=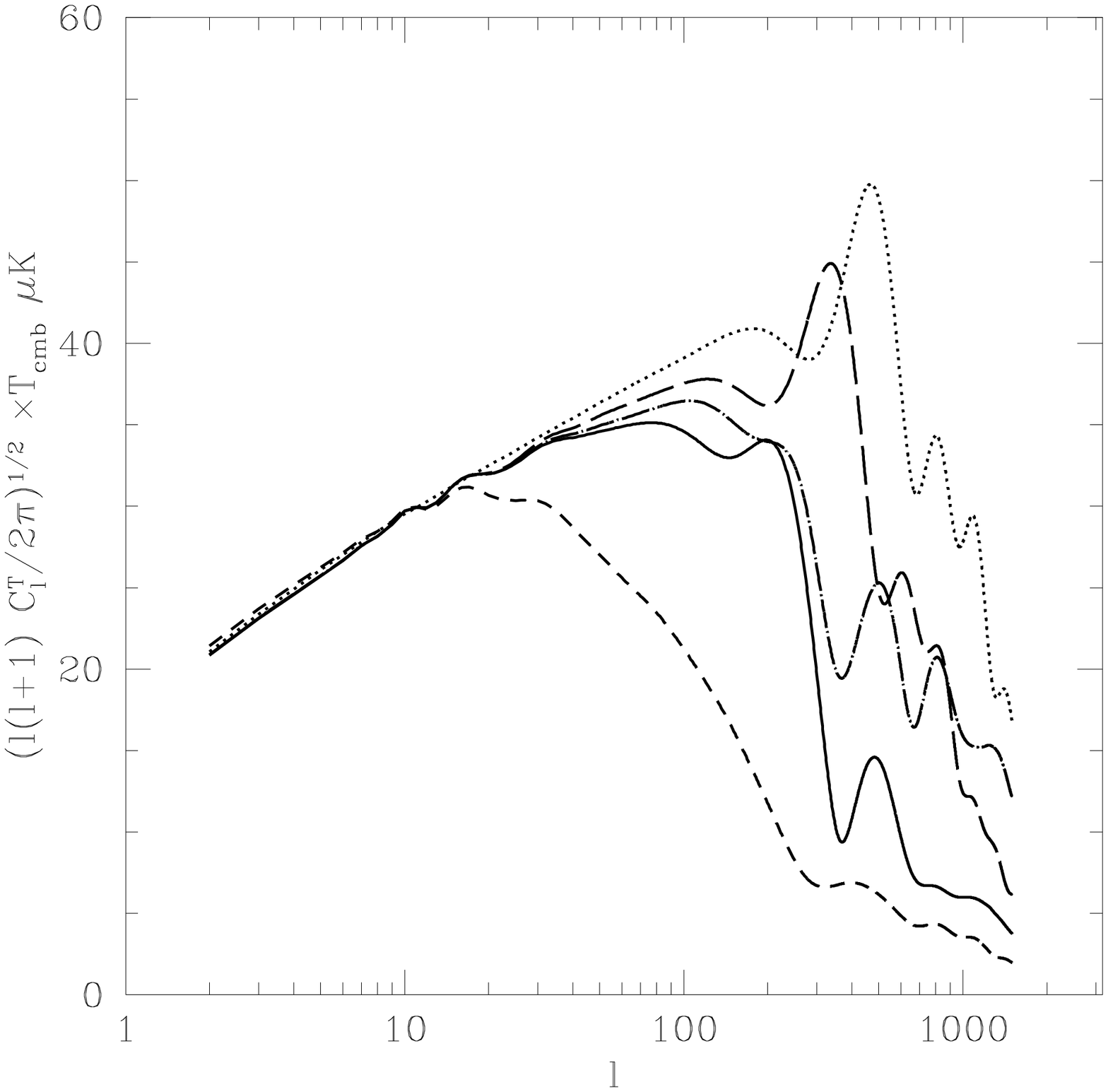,height=8cm,width=9cm}
\psfig{figure=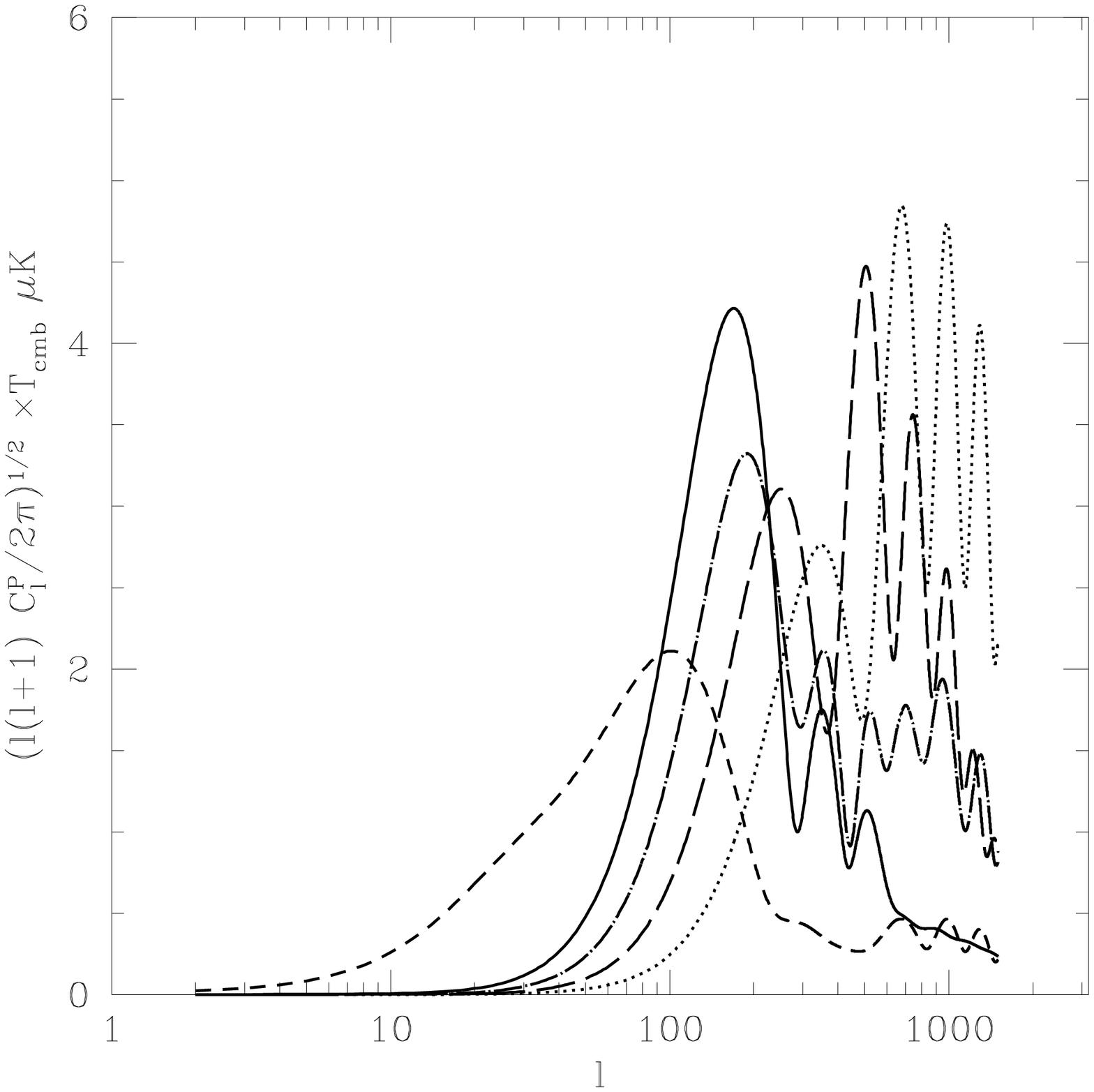,height=8cm,width=9cm}}}
\caption{\label{powerstress}Temperature (left) and polarization (right)
anisotropy power spectra for a stress source model with $A=1.0$, $B_1=1.0$ and $B_2=0.5$. The key to the curves is
the same as in fig.(\ref{powerpressure}).}
\end{figure}

Figs.~(\ref{powerpressure}) and (\ref{powerstress}) show the results of incorporating the pressure and stress sources into CMBFAST. For the pressure source the anisotropy results follow the exact pattern predicted by our analytic arguments and already confirmed in the isocurvature white noise model. Now the main contribution to the intrinsic anisotropy is from the third term of the
tight coupling solution (\ref{tight}), which moves the dominant peak in the 
temperature anisotropy to smaller scales. The delayed and shifted scenarios
comprise a single set of peaks at larger scales than in the standard case,
and in particular the main peak, which is at around $l\approx 350$ in the
standard case, is moved to around $l\approx 250$ in the delayed
scenario. In the case of {\em double last scattering} one
can still recognize the peaks from the recombination epoch on small scales.
However the expected second set of peaks from reionization is hardly
recognizeable. The polarization in each of these cases also follows the pattern already established.

One should note that in these truly active models the large-scale anisotropy is created by the ISW effect due to the existence of the sources along the line of sight. In the delayed, shifted and double last scattering scenarios reionization takes place very much before $z\approx 100$, whereas most of the ISW effect comes from sources present after this time. Therefore, reionization has very little impact on this contribution, that is, the penultimate term in (\ref{2LSS}) is effectively zero. In the total ionization scenario, where the universe is ionized for most of the time between standard last scattering and the present day, this is not necessarily the case and reionization can interfere with the anisotropy on large scales. This is illustrated in fig.~\ref{powerpressure}, although superficially it appears that the large-angle contribution has increased. This is not in fact the case since the normalization to COBE makes all the models equally around $l=10$. What has happened is a redistribution of the large-scale power and a reduction of the contribution to the COBE normalization from scales smaller than $l\approx 10$. This will not always be the case since it depends critically on the time when most of the anisotropy is created relative to the ionization history.

The results of using the stress source are very similar to those from the
pressure source as illustrated in fig.~\ref{powerstress}. However, there is one
case which is quantitatively different. It appears that the first peak in the
polarization spectrum for the shifted scenario is higher than that for the
double last scattering scenario, where in the two cases already considered
(the isocurvature white noise and pressure models) these peaks have
almost the same  height. We believe that this subtle effect is due to the
quadrapole of the temperature anisotropy, which is the source for polarization,
being non-zero when the second tight-coupling epoch begins, although we have no
analytic reasoning for this. 
 
\subsection{Causal white noise}
One of the original motivations for studying these non-standard thermal
histories was to attempt to rectify some of the observational problems of a
class of structure formation models known as causal white noise (CWN)
models~\cite{Albrecht:98,ABRW:98}. These models were spawned out of the realization that
standard scaling models with defect motivated stress-energy components are
unable to explain the observed matter fluctuations on $100h^{-1}{\rm Mpc}$
scales in a Einstein-de-Sitter universe, unless in case where large scale
biases are acceptable. It was suggested that if the source was switched off at
some point before a critical redshift ($z_c\approx 100$), then the power on
$100h^{-1}{\rm Mpc}$ scales was exactly that observed and the excess of power
on smaller scales can be rectified by modifications to cosmological
parameters. An observationally unacceptable side-effect of this is the spectrum
of CMB anisotropies produced.  

\begin{figure}[h]
\setlength{\unitlength}{1cm}
\centerline{\hbox{\psfig{figure=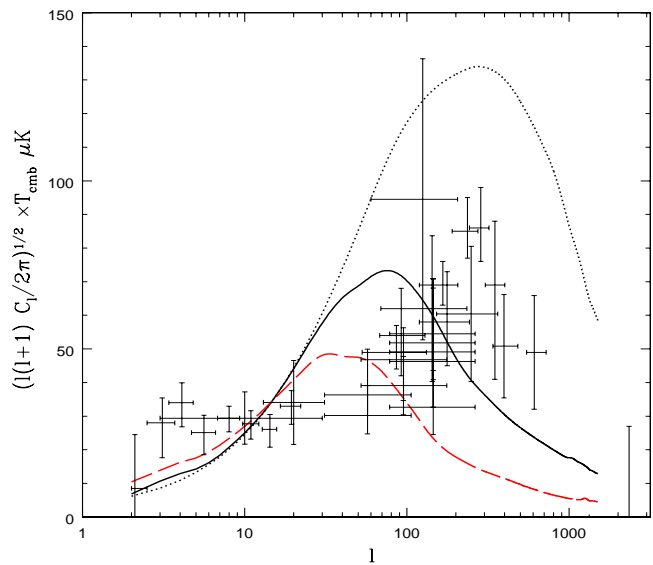,height=8cm,width=9cm}}}
\caption{\label{powercat}Temperature anisotropy power spectrum for a causal
white noise model. The dotted line is for a standard thermal history, the
solid line for {\em a} total ionization history with the parameters
$\bar{z}=400$, $\rho=350$ and $T_{\rm heat}=3\times10^7\,{\rm K}$ and the long
dashed line is for the same model with inclusion of a cosmological constant
$\Omega_\Lambda=0.65$, $\Omega_{\rm b}=0.1$ and $\Omega_{\rm c}=0.25$. The
spectra include scalar, vector and tensor contributions and are normalized
to COBE. Also included are the observations data points \protect \cite{Tegmark:data}.}
\end{figure}

Fig.~\ref{powercat} illustrates these problems for a simple CWN model in
the standard thermal history, plus the effects of our total ionization
scenario, in both an Einstein-de-Sitter cosmology and also one with a
non-zero cosmological constant. In the standard scenario we see that there
appears to be an excess of power on small scales and the shape of the
spectrum on large scales is in conflict with the COBE data. This is because
these models effectively tilt the spectrum toward smaller scales. The
action of reionization is to reduce the power on small scales to a more
acceptable level, but the problems on large-scales remain. These can be
partially relieved by the inclusion of a cosmological constant, but at the
expense of reducing the power on smaller scales. It was concluded in
ref.~\cite{Albrecht:98,ABRW:98} that although such models can explain the formation of
structure, they have considerable problems explaining the observed CMB
power spectrum. 

\subsection{Cosmic strings with a cosmological constant}
The other motivation is to
investigate whether acoustic peaks in defect models may shift to larger
angular scales in realistic thermal histories. We have already mentioned
that these models appear to predict a peak (if any at all, see
refs.~\cite{PSelTa,ABRa,ABRb,ACDKSS}) on much smaller scales than in the
standard adiabatic scenario, and in fact only very convoluted models can
rectify this~\cite{Turok:96}. One might ask is this a generic phenomena and
clearly our earlier arguments suggest that  the ionization history can be
modified to allow this to happen in more generic defect models. 

\begin{figure}
\setlength{\unitlength}{1cm}
\centerline{\hbox{\psfig{file=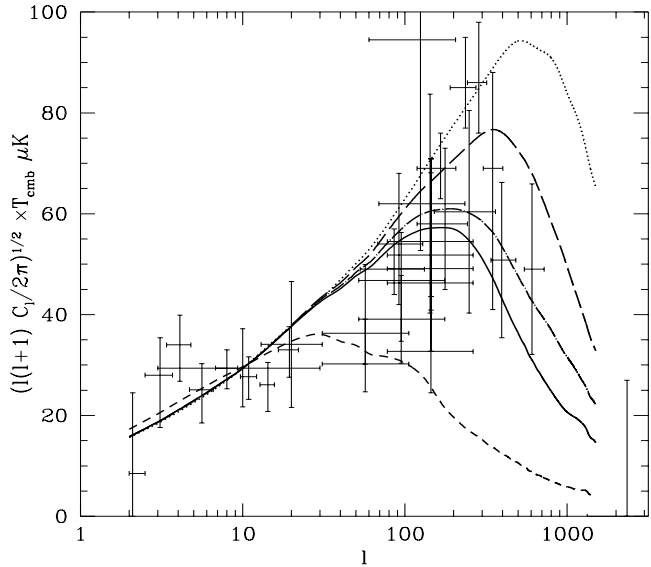,height=8cm,width=9cm}}}
\caption{The temperature anisotropy power spectra for a cosmic string
model under inclusion of a cosmological constant
$\Omega_\Lambda=0.80$. The key of the plot is the same as in fig.~\ref{powerspectrum}. We
have also included the data points with errorbars \protect \cite{Tegmark:data}.} 
\label{lambdastring}
\end{figure}

We have applied the same modified thermal histories to the model of
structure formation by cosmic strings with a cosmological constant
presented in ref.~\cite{ABRd}. The introduction of a non-zero cosmological
constant can improve the shape of the matter power spectrum and its
amplitude on scales of $100h^{-1}\,{\rm Mpc}$ to an acceptable level, and
it was shown that an important consequence of this is a broad peak in the
CMB power spectrum on around $l=500$. The model presented here has
$\Omega_{\rm B}=0.05$, $\Omega_{\rm c}=0.15$ and $\Omega_\Lambda=0.80$. In
fig.~\ref{lambdastring} we have plotted the temperature anisotropy power
spectrum for our thermal histories with this model. The spectrum behaves
very much as one might have expected from our earlier analytic arguments
and also like the simple coherent models, although it  should be noted that
in this incoherent model the concept of peaks in slightly different.  For
standard recombination (dotted line) the peak is at 
$l\approx 500$ and has an amplitude of $\approx 95 \,\mu{\rm K}$, which has
been shifted, for example, in the case of {\em double} last scattering (dot-dashed line) to a broader peak at $l\approx 200$ and a height of $60 \,\mu{\rm K}$. Similar modifications are made in the shifted and delayed last scattering scenarios. This is clearly an
improvement with respect to the current observational data, although maybe a not a necessary one.

\section{Discussion and conclusion}

We have introduced a heating source motivated by the {\em active} character of
structure formation in the context of sources like a cosmic strings. The main assumption is that there exists a phase in the universe
after recombination, where the density of active sources is large enough to
heat the baryons {\em homogenously}, at least over a short period of
time. After this the influence of the active sources on the heating
was assumed to be irrelevant for our effective description. Sources of
`late' reionization, such as, photoionization through early object
formation are not included in our analysis. We also have
not considerd the second order contribution from the Vishniac effect
\cite{Vishniac:87,Dodelson:95}, which will lead to an extra
contribution in the temperature anisotropy power spectrum at large $l$
($l>2000$). An inhomogenous treatment would also have led to such
second order contributions \cite{Hu:98}.

We calculated the spectral distortions for a wide range of heat source
parameters and established that it is hard to violate the COBE FIRAS
limits on these quantities. We extended the
framework of the semi-analytical Hu and Sugiyama formalism \cite{Hu:95a} to the
case of reionization and  analyzed four generic heat source types explaining
their influence on the CMB temperature anisotropy and polarization power
spectra for isocurvature white noise and a simple scaling source
models. The behaviour of the models was explained well by these
arguments. We have found that if there is only a shift of the surface of
last scattering, the anisotropy power spectrum becomes damped and the peaks
are shifted to larger scales. If there appears a second surface of last
scattering the acoustic peak structure changes and the suppression
on small scales is not as large as for a just shifted last scattering
surface. The most important feature of the polarization  power spectrum 
is the appearance of a prominent contribution on intermediate scales due to
reionization. We then applied the source models to CWN models and also to a
realistic cosmic string model. We found that it was possible to reduce the
amount of power on small scales in CWN models and to move the peak in
string models to larger scales --- the original motivation for this
work. However, it appears to not be possible to have a substantial effect
on the spectrum at large angles in CWN models.  

\section*{Acknowledgments}

We thank U. Seljak and M. Zaldarriaga for the use of CMBFAST and L. Knox,
R. Crittenden, M. Rees and A. Stebbins for helpful conversations. The model for cosmic strings used to produce fig.~\ref{lambdastring} was developed by AA and RAB in collaboration with James Robinson. This work was
supported by PPARC and some of the computations were done at the UK National Cosmology Supercomputing Center, funded by PPARC, HEFCE and Silicon Graphics/Cray Research. RAB is funded  by Trinity College. JW is supported by a
DAAD fellowship HSP III financed by the German Federal Ministry for Research
and Technology.  

\def\jnl#1#2#3#4#5#6{\hang{#1, {\it #4\/} {\bf #5}, #6 (#2).}}
\def\jnltwo#1#2#3#4#5#6#7#8{\hang{#1, {\it #4\/} {\bf #5}, #6; {\it
ibid} {\bf #7} #8 (#2).}} 
\def\prep#1#2#3#4{\hang{#1, #4.}} 
\def\proc#1#2#3#4#5#6{{#1 [#2], in {\it #4\/}, #5, eds.\ (#6).}}
\def\book#1#2#3#4{\hang{#1, {\it #3\/} (#4, #2).}}
\def\jnlerr#1#2#3#4#5#6#7#8{\hang{#1 [#2], {\it #4\/} {\bf #5}, #6.
{Erratum:} {\it #4\/} {\bf #7}, #8.}}
\def\prl{Phys.\ Rev.\ Lett.}
\def\pr{Phys.\ Rev.}
\def\pl{Phys.\ Lett.}
\def\np{Nucl.\ Phys.}
\def\prp{Phys.\ Rep.}
\def\rmp{Rev.\ Mod.\ Phys.}
\def\cmp{Comm.\ Math.\ Phys.}
\def\mpl{Mod.\ Phys.\ Lett.}
\def\apj{Ap.\ J.}
\def\apjl{Ap.\ J.\ Lett.}
\def\aap{Astron.\ Ap.}
\def\cqg{Class.\ Quant.\ Grav.} 
\def\grg{Gen.\ Rel.\ Grav.}
\def\mn{MNRAS}
\def\ptp{Prog.\ Theor.\ Phys.}
\def\jetp{Sov.\ Phys.\ JETP}
\def\jetpl{JETP Lett.}
\def\jmp{J.\ Math.\ Phys.}
\def\zpc{Z.\ Phys.\ C}
\def\cupress{Cambridge University Press}
\def\pup{Princeton University Press}
\def\wss{World Scientific, Singapore}
\def\oup{Oxford University Press}

\end{document}